\documentclass{aa}

\usepackage{graphicx}
\usepackage{subcaption}
\usepackage{color}

\usepackage{xspace}
\newcommand{\psr}{PSR\,J0537$-$6910\xspace}

\usepackage{txfonts}\usepackage[colorlinks=true,linkcolor=blue,filecolor=blue,citecolor=blue,urlcolor=blue]{hyperref}
\begin{document}

   \title{PSR J0537--6910: Exponential recoveries detected for 12 glitches}

   \author{E. Zubieta
          \inst{1,2}\fnmsep\thanks{E-mail: ezubieta@iar.unlp.edu.ar}
          \and
          C. M. Espinoza\inst{3,4}
          \and
          D. Antonopoulou\inst{5}
          \and
          W. C. G. Ho\inst{6}
          \and
          L. Kuiper\inst{7}
          \and
          F. Garc\'ia\inst{1,2}
          \and
          S. del Palacio\inst{1,8}
          }

   \institute{Instituto Argentino de Radioastronom\'ia (CCT La Plata, CONICET; CICPBA; UNLP),
              C.C.5, (1894) Villa Elisa, Buenos Aires, Argentina
         \and
             Facultad de Ciencias Astron\'omicas y Geof\'{\i}sicas, Universidad Nacional de La Plata, Paseo del Bosque, B1900FWA La Plata, Argentina
         \and 
            Departamento de F\'isica, Universidad de Santiago de Chile (USACH),  Av. V\'ictor Jara 3493, Estaci\'on Central, Chile.
         \and 
            Center for Interdisciplinary Research in Astrophysics and Space Sciences (CIRAS), Universidad de Santiago de Chile.
        \and
            Jodrell Bank Centre for Astrophysics, Department of Physics and Astronomy, The University of Manchester, Manchester M13 9PL, UK
        \and
             Department of Physics and Astronomy, Haverford College, 370 Lancaster Avenue, Haverford, PA 19041, USA
        \and
            SRON - Space Research Organisation Netherlands, Niels Bohrweg 4, 2333 CA, Leiden, The Netherlands
        \and
            Department of Space, Earth and Environment, Chalmers University of Technology, SE-412 96 Gothenburg, Sweden
            }
   \date{Received ...; accepted ...}
 
  \abstract
  {Pulsar rotational glitches are unresolved increments of the rotation rate that sometimes trigger an enhancement of the spin-down rate. 
  On occasions, the augmented spin-down decays gradually over days or months in an exponential manner. 
  This is observed particularly after the largest glitch events. 
  Glitches and their exponential recoveries are attributed to the presence of a neutron superfluid inside pulsars.
  The young pulsar \psr exhibits the highest known glitching rate, with 60 events detected in nearly 18 years of monitoring. }
  {Despite most \psr glitches being large, only one exponential recovery has been reported, following its first discovered glitch. This is puzzling as pulsars of similar age, rotational properties, and glitching behaviour, typically present significant exponential recoveries.  
  We wish to understand if this is an intrinsic difference of \psr or a detectability issue due, for example, to its high glitch frequency.
  }
  {The full dataset, including recent NICER observations, was systematically searched for the signal of exponential relaxations. 
  Each glitch was first tested for evidence of a possible recovery on a broad range of trial timescales.
  Good candidates were investigated further, by finding the best recovery models with and without an exponential term, and comparing them using Bayesian evidence.
  } 
  {We discovered 6 new glitches, bringing the total number of glitches in this pulsar to 66. Our search and selection criteria strongly indicate the presence of further 11, previously undetected, exponential recoveries. We provide updated glitch and timing solutions. Exponential recoveries were detected only for the largest glitches ($\Delta\nu>20\,\mu$Hz), though not for all of them. 
  The inferred exponential timescales vary between $4$ and $37$\,d, with the decaying frequency increment being close to or below 1\% of the total in general.    
  {We find that the second derivative of the spin frequency ($\ddot\nu$) can remain fairly stable across several glitches, and only some glitches are associated with  persistent $\ddot{\nu}$ changes. In particular, $\ddot\nu$ tends to take the lowest values after glitches with exponential recoveries, corresponding to inter-glitch braking indices $n_{\rm{ig}} \sim 6 - 9$. On the other hand, following glitches without exponential recoveries, even large ones, $\ddot{\nu}$ tends to be higher, with varied values that typically lead to braking indices above 9 (with inferred $n_{\rm{ig}}$ mostly clustered between 10 and 35).  }
  }
   {}

   \keywords{(Stars:) pulsars: general -- 
                 methods: observational  --
                pulsars: individual: PSR J0537--6910
               }

   \maketitle

\section{Introduction}

Pulsars are rapidly rotating neutron stars that emit radiation across multiple wavelengths, from radio to gamma rays. They have extremely high and stable moments of inertia, which render them with remarkably steady rotations. This places them among the most accurate clocks in the Universe \citep{2012MNRAS.427.2780H}. 

However, the smooth rotation of pulsars can sometimes be interrupted by sudden increases of the rotational frequency $\nu$, called glitches \citep{2011MNRAS.414.1679E,2018IAUS..337..197M, 2023MNRAS.521.4504Z}.
These spin-up episodes often also present a sudden change in the pulsar spin-down rate, which then recovers, sometimes patrialy in an exponential-like manner \citep{2013MNRAS.429..688Y,2024MNRAS.532..859L}. 
Currently, around 700 glitches have been detected in more than 200 pulsars\footnote{http://www.jb.man.ac.uk/pulsar/glitches.html}, with a great range of amplitudes: observed frequency changes $\Delta\nu$ vary from $\sim10^{-4}\;\rm{\mu Hz}$ to nearly $100\;\rm{\mu Hz}$ \citep{2022MNRAS.510.4049B}.

Glitches are believed to be the manifestation of a sudden angular momentum transfer between a loosely coupled superfluid interior and the rest of the star \citep{1969Natur.224..673B,1975Natur.256...25A,2015IJMPD..2430008H, 2022RPPh...85l6901A,2022Univ....8..641Z}. The neutrons in the inner crust and core of neutron stars are in a superfluid state, which is characterised by irrotational flow with vanishing friction. Inside the fast-rotating pulsars it is, however, energetically favourable for the superfluid to follow its normal surroundings' rotation whilst remaining irrotational in its bulk. It does so by forming many microscopic vortices with non-superfluid cores, which carry all circulation and whose density determines a local, average, rotation rate. As the pulsar spins down, vortices move outwards and eventually vanish at the superfluid boundary so that the superfluid decelerates as well. If in some stellar layer this vortex motion is prevented by forces that `pin' vortices to the normal matter, then excess vorticity builds up because the pulsar spins down whilst the local superfluid maintains a higher rotation rate. Eventually, either due to the increasing velocity difference between the components, or also aided by some external disturbance, a catastrophic unpinning event takes place, and vortices move, quickly transferring the excess angular momentum to the crust that spins-up at a glitch. 

Weakly coupled superfluid regions lag behind the abrupt glitch-induced increase of the crust's rotation rate and temporarily decouple, reducing the effective moment of inertia upon which the external, spin-down, torque acts. This causes a transient increase in the spin-down rate $|\dot{\nu}|$, commonly observed to accompany glitches, with typical amplitudes up to $\Delta\dot{\nu}\sim0.1\%\;\dot{\nu}$. In a considerable minority, mostly consisting of large glitches in young pulsars, part of this change exponentially decays over characteristic relaxation timescales of days to several weeks: a signature of superfluid components slowly recoupling to the crust \citep{2015IJMPD..2430008H,2020MNRAS.499..161H, 2022RPPh...85l6901A}. For well-monitored neutron stars with regular large glitches like the Vela pulsar, detected decaying timescales range from minutes 
\citep{2002ApJ...564L..85D, 2018Natur.556..219P,2019NatAs...3.1143A,2025A&A...698A..72Z} to $\sim500~\mathrm{d}$ \citep{2024A&A...689A.191Z}. The study of glitches, and particularly their relaxation process, is key to understanding the internal physics of neutron stars.

\psr stands out as the fastest spinning ($\nu\sim62\,$Hz), non-recycled, rotation-powered pulsar with the highest known spin-down energy loss rate 
($\dot{E} \sim 4.88 \times 10^{38}\,\mathrm{erg}\,\mathrm{s^{-1}}$). 
It is a young object, associated with the 1--5~kyr old supernova remnant N157B in the Large Magellanic Cloud \citep{1998ApJ...494..623W,1998ApJ...499L.179M}, and it is by far the most frequently glitching pulsar with a rate of approximately 3 events per year and a total of 60 reported glitches in nearly 18 years of cumulated data \citep{2006ApJ...652.1531M,2018MNRAS.473.1644A,2018ApJ...852..123F,2022ApJ...939....7H}.
About 70\% of these glitches are large, with $\Delta\nu>10\,\mu$Hz (90\% above $1\,\mu$Hz), which contribute to the peak at large values observed in the bimodal distribution of glitch sizes for all glitching pulsars \citep{2017A&A...608A.131F}. 

Glitches in \psr are accompanied by significant increases in the spin-down rate $|\dot{\nu}|$, which dominate the $\dot{\nu}$-time evolution (Fig. \ref{fig: spin-down}); a common feature of glitching pulsars of similar age, like Vela \citep{2015MNRAS.446..857L,2017MNRAS.466..147E}.
On shorter terms, these pulsars exhibit inter-glitch $\ddot{\nu}$ values considerably larger than the predictions of any model for pulsar braking. 
For example, the dipole braking model, in which the neutron star loses energy exclusively by magnetic dipole radiation, assuming constant moment of inertia and magnetic field, predicts a $\ddot\nu$ rate that is 20 times smaller than the observed values between glitches in \psr.
Contrary to similar (glitching) pulsars, however, \psr does not feature prominent exponential relaxations and their presence has been somewhat underexplored \citep{2020MNRAS.498.4605H}. 
The first glitch observed was the largest and the only one for which an exponential recovery was clearly discernible and well-studied \citep{2018MNRAS.473.1644A}. 

In this work, we present 6 new glitches detected in recent NICER (\textit{Neutron star Interior Composition Explorer}) observations of \psr, and analyze all known glitches in the entire dataset for the presence of exponential recoveries. 
In Sect.~\ref{sec: obs}, we summarize the reduction process of the X-ray data from NICER and RXTE (\textit{Rossi X-ray Timing Explorer}). In Sect.~\ref{sec: methods}, we develop a systematic methodology to identify and characterize post-glitch exponential recoveries. In Sect.~\ref{sec: results}, we report on the new 6 glitches and the results of the search for exponential recoveries on the total of 66 glitches observed so far. Finally, in Sect.~\ref{sec: discussion}, we discuss the implications of our findings.

\section{Observations}\label{sec: obs}

\psr was discovered in the X-ray band in 1998 \citep{1998ApJ...499L.179M} and only recently some weak and dispersed pulses in the radio were detected \citep{2005AdSpR..35.1181C,2024AAS...24313204C}. 
Monitoring of its rotation with RXTE began on 1999 January 19 and lasted until 2011 December 31.
Regular observations resumed $5.45$ years later with NICER \citep{2020MNRAS.498.4605H}.
In this work, we use all available RXTE and NICER data for this pulsar and calculate pulse times of arrival (TOAs) by cross-correlating pulse profiles with a pulse template.

The RXTE TOAs used in our analysis are the same ones as those used in \cite{2018MNRAS.473.1644A}, which were derived using a method similar to the one presented by \cite{2009A&A...501.1031K} for PSR\,J1846$-$0258. 

However, the template used for the correlation with the folded profiles to obtain a TOA measurement comes from a fit of a 60-bin asymmetric Lorentzian to a high signal-to-noise profile. In some cases, consecutive observations from RXTE were combined to achieve sufficient exposure to measure the spin frequency of the pulsar confidently. 
This happened mainly with observations performed after 2006, where generally only one or two of the five proportional counter units (PCUs) were operational during a standard observation.

For the NICER data, the methodology was laid out in \cite{2020MNRAS.498.4605H}: \textsc{HEASoft 6.22–6.35.1} \citep{2014ascl.soft08004N} and \textsc{NICERDAS 2018-03-01\_V003-2025-
03-11\_V013A} were used to process and filter the observations. We kept only events in the 1--7~keV range, where the pulse is easily detected \citep{2015MNRAS.449.3827K}. As for RXTE data, there were some cases for which sets of close individual observations had to be merged to significantly detect the pulsar period. The typical total exposure of merged observations ranged between 4~and~9~ks. From each merged observation, one TOA was obtained. In this case, TOAs were measured by correlating the folded profiles with a template that we obtained from fitting a Gaussian to a series of NICER high signal-to-noise pulse profiles.

\section{Methods}\label{sec: methods}

In the following, we use the 45 glitches as published by \citet[][]{2018MNRAS.473.1644A} for the complete RXTE observing period (MJD 51197 to 55926),  
and 15
reported glitches from NICER observations, which we enumerate from 46 onward: timing solutions for glitches 46 to 53 can be found in \cite{2020MNRAS.498.4605H} (glitches 1 to 7 in their work),  glitches 54 to 56 were presented in \cite{2021ApJ...913L..27A} (glitches 8 to 10 in their work), and glitches 57 to 60 in  \cite{2022ApJ...939....7H} (11 to 15 in their work). 

Additionally, we processed the latest NICER data (after MJD 59627) and report on six new glitches that we identified and measured. Therefore a total of 66 glitches of \psr were included in the post-glitch recovery analysis.

\subsection{Pulsar timing}

The rotation of pulsars is monitored by recording TOAs of detected pulses and comparing them to the predictions of a spin-down model for the rotation phase $\phi$, which can be approximated by a truncated Taylor series:

\begin{equation}\label{eq:timing-model}
    \phi(t)=\phi+\nu_0(t-t_0)+\frac{1}{2}\dot{\nu}_0(t-t_0)^2+\frac{1}{6}\Ddot{\nu}_0(t-t_0)^3 \quad, 
\end{equation}

\noindent where $\nu_0$, $\dot\nu_0,$ and $\ddot\nu_0$ are the rotation frequency of the pulsar and its first and second derivatives, respectively, at the reference epoch $t_0$. 
The differences between the observed and predicted arrival times are called timing residuals. These can be remarkably small, thanks to the accuracy of TOAs measurements and the stability of the pulsar's rotation.
Glitches introduce well-known signals in the residuals and, unless they are too small, are generally straightforward to identify.

To describe the changes in the rotation of the pulsar due to a glitch at time $t_\mathrm{g}$ and its post-glitch recovery, an additional phase $\phi_\mathrm{g}$ must be added to the model of Eq.~\ref{eq:timing-model} for  $t>t_\mathrm{g}$, of the form \citep[e.g.][]{2013MNRAS.429..688Y,2025A&A...694A.124Z}:

\begin{multline}\label{eq:glitch-model}
    \phi_\mathrm{g}(t) = \Delta \phi + \Delta \nu_\mathrm{p} (t-t_\mathrm{g}) + \frac{1}{2} \Delta \dot{\nu}_\mathrm{p} (t-t_\mathrm{g})^2 + \\ 
    \frac{1}{6} \Delta \Ddot{\nu}_\mathrm{p}(t-t_\mathrm{g})^3 + \left[1-\exp{\left(-\frac{t-t_\mathrm{g}}{\tau_\mathrm{d}}\right)} \right]\Delta \nu_\mathrm{d} \, \tau_\mathrm{d} \;\text{.}
\end{multline}

\noindent The quantities $\Delta \nu_\mathrm{p}$, $\Delta \dot{\nu}_\mathrm{p}$, and $\Delta \Ddot{\nu}_\mathrm{p}$ describe persisting changes of the rotational parameters with respect to the pre-glitch model, whilst a non-zero $\Delta \phi$ might be required due to the uncertainty of the precise glitch epoch $t_\mathrm{g}$. The last term describes a transient component of the spin-up, $\Delta \nu_\mathrm{d}$, that exponentially relaxes on a timescale $\tau_\mathrm{d}$. Glitches can have one or more such exponential components in their recoveries; however, in this study of \psr we examine glitch models with a maximum of one exponential term.  
We compute the total, unresolved, initial frequency step as $\Delta \nu =   \Delta \nu_\mathrm{p} + \Delta \nu_\mathrm{d}$ and its recovery factor as $Q=\Delta\nu_\mathrm{d} / \Delta\nu$. The equivalent change in frequency derivative is $\Delta \dot{\nu} = \Delta \dot{\nu}_\mathrm{p} - \left( \Delta \nu_\mathrm{d}/\tau_\mathrm{d}\right) $, and it is useful to also define the spin-down rate recovery factor as $Q'=\Delta\nu_\mathrm{d}/\left(\Delta\nu_\mathrm{d}-\tau_\mathrm{d} \Delta\dot\nu_\mathrm{p}\right)$ \citep{2011MNRAS.411.1917W}.

\subsection{Post-glitch recovery searches}\label{sec:methodology}

We analyzed each glitch ($n$) using the TOAs between the previous glitch ($n-1$) and the following one ($n+1$). Within this data span, we used {\sc tempo2} to fit Eq. \ref{eq:timing-model} to the pre-glitch TOAs. If an exponential recovery was detected for glitch $n-1$, we excluded any TOAs that were clearly affected by it from the pre-glitch dataspan used to characterize the following glitch $n$. 
This ensures that the pre-glitch-$n$ data can be accurately described just by Eq.~\ref{eq:timing-model}. 
In general, \psr has a high inter-glitch $\Ddot \nu$, which can be determined even for short time intervals. We therefore included this term in the fits except in the cases of exceptionally short (four or fewer TOAs) pre-glitch datasets, for which we set $\Ddot \nu=1\times10^{-20}~\mathrm{Hz~s^{-2}}$ \citep{2020MNRAS.498.4605H}.

Indications of exponential signatures in the post-glitch recovery were searched for automatically in each inter-glitch dataset comprising of at least six TOAs. 
The first step of our methodology was to make an initial estimate of the glitch size by fitting a glitch model with only $\Delta \phi$, $\Delta \nu_\mathrm{p}$, and $\Delta \dot{\nu}_\mathrm{p}$ (keeping $\Delta\nu_\mathrm{d}=\Delta\Ddot\nu_\mathrm{p}=0$). 
Glitch parameters from the literature were used as a starting solution, when available. 
In the next step, the full glitch model of Eq.~\ref{eq:glitch-model} was fitted for, with $\Delta \nu_\mathrm{d}$ and all other parameters allowed to vary except for $\tau_\mathrm{d}$, which was set at a trial value. 
For each post-glitch dataset, we performed such fits for a range of possible relaxation timescales: all integer values between one day and four times the total length of the post-glitch data span were used as trials for $\tau_\mathrm{d}$. Exponential recoveries on shorter timescales cannot be accurately determined due to the timing resolution of the dataset. Similarly, the exponential signature is hard to discern if the relaxation timescale is considerably longer than the post-glitch fitted interval. 

\begin{figure}%[h!]
    \includegraphics[width=\linewidth]{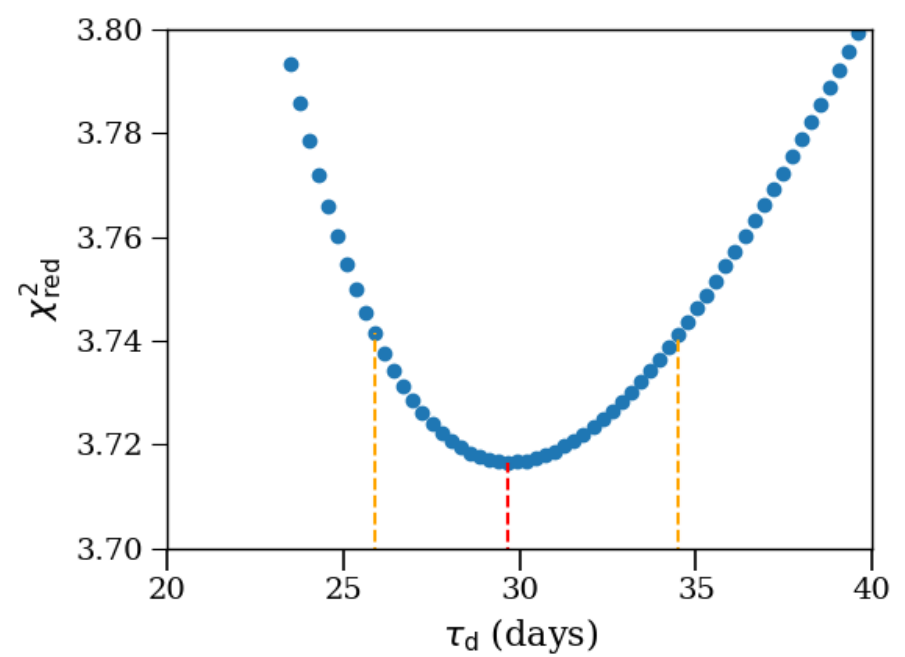}
    \caption{Example of the exploratory process to determine possible values for an exponential relaxation timescale. The blue dots corresponds to the $\chi^2_\mathrm{red}$ values of glitch \#64, for a subset of the total range of trial $\tau_\mathrm{d}$ values, focused around the timescale that returns the minimum $\chi^2_\mathrm{red}$ (marked by the red line).  Orange dashed lines correspond to $1\sigma$ uncertainty. These, rather large $\chi^2_\mathrm{red}$ values are only used to identify a possible timescale. Further analyses are performed for the final measurements (Table \ref{tab:glitch_solutions}).} 
    \label{fig: tau_sist}
\end{figure}

The variations of the $\chi^2_\mathrm{red}$ metric for each glitch fit with respect to $\tau_\mathrm{d}$ were analyzed. When a minimum was present, the search was refined around it, as exemplified in Fig.~\ref{fig: tau_sist}. 

This is not the final determination of the glitch recovery but rather a preliminary exploration to identify whether the data favours a possible recovery timescale. Glitches for which the minimum $\chi^2_\mathrm{red}$ occurred at either the beginning or the end of the explored $\tau_\mathrm{d}$ range were not further explored: we assume in such cases that the absolute minimum might occur outside our trial $\tau_\mathrm{d}$ range and therefore, as discussed above, we cannot securely detect and characterize an exponential recovery with the available data. 

For the final step, we considered two different glitch models to compare through marginal likelihood:
\begin{itemize}
    \item Model 1 (without recovery): Consists of $\Delta \phi$, $\Delta \nu_\mathrm{p}$ and $\Delta \dot{\nu}_\mathrm{p}$. We also considered $\Delta \Ddot{\nu}_\mathrm{p}$ when its inclusion led to a significant decrease of the residuals' $\mathrm{rms}$. 
    \item Model 2 (with one exponential recovery): Consists of $\Delta \phi$, $\Delta \nu_\mathrm{p}$, $\Delta \dot{\nu}_\mathrm{p}$, $\Delta \nu_\mathrm{d}$ and $\tau_\mathrm{d}$. $\Delta \Ddot{\nu}_\mathrm{p}$ was also included when its inclusion led to a significant decrease of the residuals' $\mathrm{rms}$. We used the best-fit recovery parameters from the previous step as initial values for these fits.
\end{itemize}
A nested sampling code\footnote{\url{https://github.com/kbarbary/nestle}} \citep{2004AIPC..735..395S} based on the PINT software \citep{2021ApJ...911...45L} was used to fit the models and infer the most likely values for the glitch parameters. We explored a wide range for each parameter to ensure that a global solution was reached. 

\renewcommand{\arraystretch}{1.25}

\begin{table*}[]
    \setlength{\tabcolsep}{4pt}
    \centering
    \caption{Timing solutions (without exponential recovery terms) for the six new glitches detected with NICER. }
    \begin{tabular}{cccccccccc}
        \hline\hline
        Glitch  & $t_0$  & $t_g$  & $\nu$ & $\dot\nu$ & $\Ddot\nu$ & $\Delta\phi$ & $\Delta\nu_p$ & $\Delta\dot\nu_p$ & $\Delta\Ddot{\nu}_\mathrm{p}$  \\ 
          &  (MJD) & (MJD) & $(\mathrm{Hz})$ & ($10^{-10}~\mathrm{Hz~s^{-1}})$ & ($10^{-20}~\mathrm{Hz}$ $\mathrm{s^{-2}}$) & (cycle) & ($\mu\mathrm{Hz}$) &  ($10^{-13}~\mathrm{Hz}$) &($10^{-20}~\mathrm{Hz}~\mathrm{s^{-2}}$)  \\ 
        \hline
        61 & 59607 & 59692(6) & 61.8972423813(5) &  $-$1.997915(3) & 0.80(8) &  $-$0.18(2) & 5.83(1) & $-$0.86(6) & --- \\ 
        62 & 59716 & 59741(6) & 61.895366836(4) & $-$1.99802(6) & 1$\dagger$ &  $-$0.42(2) & 15.40(2) & $-$1.06(7) & 0.56(4) \\
        63 & 59813 & 59886(5) &  61.893707533(1) & $-$1.997897(5) & 1.56(2) &  0.10(1) & 24.375(9) & $-$1.77(4) & $-$0.71(5) \\ 
        64 & 59958 & 60028(2) & 61.891228926(7) & $-$1.998156(5) & 0.84(2) &  1.89(3) & 34.22(2) & $-$1.43(6) & -2(6) \\
        65 & 60150 & 60222(2) & 61.887948098(4) & $-$1.998211(2) & 0.58(2) &  0.223(7) & 21.105(9) & $-$7.22(9) & -- \\
        66 & 60290 & 60379(22) &  61.885552197(1) & $-$1.998404(5) & 0.90(2) &  0.39(6) & 31.49(3) & $-$1.22(9) & $-$1.3(8) \\
         \hline
    \end{tabular}
    \label{tab:new_solutions}
    \tablefoot{Solutions and their uncertainties were obtained by fitting the TOAs using {\sc tempo2}.
    Glitch numbers follow the counting of RXTE glitches \citep{2018MNRAS.473.1644A}. To recover the NICER count, subtract 45. $\dagger$ indicates that $\Ddot\nu$ was fixed at that value. The numbers in parenthesis are the uncertainties in the last digit.}
\end{table*}

Finally, to decide the appropriate model for each glitch, we calculated their Bayesian evidences, $Z_1$ and $Z_2$, for Models 1 and 2, respectively, and assessed their logarithmic ratio: 
\begin{equation}
    \ln{B_{21}} = \ln{Z_2} - \ln{Z_1},
\end{equation}
\noindent following the criterion used in \citet{2024MNRAS.532..859L}: if $\mathrm{ln} B_{21}<2.5$, Model 2 is considered to have insufficient evidence over Model 1. In the case of $2.5<\mathrm{ln} B_{21}<5$, Model 2 is considered to have moderate evidence over Model 1. For $5<\mathrm{ln} B_{21}<10$, strong evidence for Model 2 over Model 1 is considered, while if $\mathrm{ln} B_{21}>10$, evidence for Model 2 is considered as decisive.

We treated all glitches for which the evidence for Model 2 was moderate, strong, or decisive as detections of exponential relaxations. 
In some cases, the posterior distributions for certain parameters are considerably asymmetric, and therefore we use the posterior's mode -- instead of the mean value -- as representative for each parameter. From these, we constructed a timing solution for each glitch and used it as a starting solution for a final fit, which was obtained using PINT.
In most cases, the final fit only marginally reduced the RMS of the residuals compared to the starting solution.
The difference between the modes and the values obtained after this fit was always smaller than the range defined by the $68\%$ of the most likely values in the posterior distributions, even for the most asymmetric posteriors. 
We assign each final parameter uncertainties calculated as its distance to the 16th and 84th percentiles of its posterior samples.

\renewcommand{\arraystretch}{1.15}

\begin{table*}[h]
    \centering
    \caption{Solutions for glitches that presented an exponential recovery.}
    \begin{tabular}{cccccccccc}
        \hline\hline
        Glitch & $t_g$ & $\Delta \phi$ & $\Delta\nu_\mathrm{p}$ & $\Delta \dot \nu_\mathrm{p}$ & $\Delta\Ddot{\nu}_\mathrm{p}$  & $\tau_\mathrm{d}$ & $\Delta\nu_\mathrm{d}$ & $Q$ & $Q'$  \\ 
        & (MJD) & (cycle) & ($\mu\mathrm{Hz}$) & ($10^{-13}~\mathrm{Hz}$ $\mathrm{s^{-1}}$) & ($10^{-20}~\mathrm{Hz}$ $\mathrm{s^{-2}}$) & (days) & ($\mu\mathrm{Hz}$) & (\%) & (\%) \\
        \hline
        \noalign{\vskip 3pt}
        1 & 51278(16) & $0.5^{+0.1}_{-0.1}$ & $42.27^{+0.02}_{-0.02}$ & $-0.70^{+0.04}_{-0.04}$ & $-$ & $21^{+5}_{-5}$ & $0.3_{-0.1}^{+0.1}$ & $0.8_{-0.2}^{+0.2}$ & $71_{-8}^{+8}$  \\ 
        \noalign{\vskip 3pt}
        2 & 51562(15) & $0.6^{+0.1}_{-0.1}$ & $27.8_{-0.1}^{+0.1}$ & $-1.2_{-0.1}^{+0.1}$ & $-$ & $31_{-12}^{+28}$ & $0.3_{-0.1}^{+0.1}$ & $1.0_{-0.4}^{+0.5}$ & $46_{-24}^{+16}$ \\ 
        \noalign{\vskip 3pt}
        6 & 51960(5) & $0.0_{-0.2}^{+0.1}$ & $28.14_{-0.02}^{+0.03}$ & $-1.5_{-0.1}^{+0.1}$ & $-2.2_{-0.3}^{+0.3}$ & $8_{-5}^{+2}$ & $0.2_{-0.1}^{+0.6}$ & $0.8_{-0.4}^{+2.2}$ & $69_{-12}^{+61}$ \\ 
        \noalign{\vskip 3pt}
        12 & 52545(6) & $0.57_{-0.06}^{+0.06}$ & $26.09^{+0.01}_{-0.01}$ & $-0.73^{+0.03}_{-0.04}$ & $-$ & $7_{-5}^{+7}$ & $0.05_{-0.06}^{+0.07}$ & $0.2_{-0.2}^{+0.2}$ & $97_{-7}^{+2}$ \\ 
        \noalign{\vskip 3pt}
        38 & 55280(4) & $0.5_{-0.2}^{+0.1}$ & $34.02^{+0.02}_{-0.02}$ & $-0.4^{+0.1}_{-0.1}$ & $-$ & $4_{-1}^{+4}$ & $0.8_{-0.5}^{+1.1}$ & $2_{-1}^{+3}$ & $98_{-1}^{+2}$  \\ 
        \noalign{\vskip 3pt}
        43 & 55615(4) & $0.1_{-0.1}^{+0.1}$ & $28.0_{-0.1}^{+0.1}$ & $-$ & $-$ & $18_{-8}^{+14}$ & $0.2_{-0.1}^{+0.1}$ & $0.8_{-0.3}^{+0.4}$ & $100_{-12}^{+7}$ \\ 
        \noalign{\vskip 3pt}
        45 & 55819(2) & $-0.2_{-0.1}^{+0.1}$ & $20.7_{-0.5}^{+0.3}$ & $-$ & $-$ & $37_{-10}^{+30}$ & $0.8_{-0.2}^{+0.5}$ & $4_{-1}^{+2}$ &  $100_{-21}^{+7}$  \\ 
        \noalign{\vskip 3pt}
        47 & 58152(11) & $0.2_{-0.2}^{+0.4}$ & $36.02_{-0.09}^{+0.07}$ & $-1.5^{+0.3}_{-0.3}$ & $-$ & $6_{-2}^{+7}$ & $0.5_{-0.7}^{+0.6}$ & $1_{-2}^{+2}$ & $87_{-24}^{+26}$ \\ 
        \noalign{\vskip 3pt}
        53 & 58868(5) & $-0.0_{-0.1}^{+0.2}$ & $23.96_{-0.09}^{+0.07}$ & $-2.1^{+0.4}_{-0.3}$ & $-5.0^{+0.6}_{-0.7}$ & $5_{-1}^{+5}$ & $0.5_{-0.4}^{+0.1}$ & $2.0_{-0.8}^{+0.8}$ & $84_{-23}^{+36}$ \\ 
        \noalign{\vskip 3pt}
        56 & 59103(6) & $0.4_{-0.1}^{+0.1}$ & $33.81^{+0.02}_{-0.01}$ & $-1.3^{+0.2}_{-0.2}$ & $-3.1^{+0.2}_{-0.2}$ & $9_{-3}^{+5}$ & $0.2^{+0.4}_{-0.1}$ & $0.6^{+1.1}_{-0.3}$ & $37^{+34}_{-20}$\\ 
        \noalign{\vskip 3pt}
        64 & 60028(1) & $0.83_{-0.04}^{+0.07}$ & $34.0_{-0.1}^{+0.1}$ & $-1.0_{-0.1}^{+0.2}$ & $-0.4^{+0.1}_{-0.1}$ & $30_{-8}^{+13}$ & $0.4_{-0.1}^{+0.1}$ & $1.1_{-0.3}^{+0.4}$& $61_{-12}^{+11}$  \\ 
        \noalign{\vskip 3pt}
        65 & 60223(3) & $0.22_{-0.04}^{+0.05}$ & $21.1_{-0.1}^{+0.1}$ & $-0.7_{-0.1}^{+0.1}$ & $-$ & $33_{-11}^{+18}$ & $0.3_{-0.1}^{0.1}$ & $1.3_{-0.4}^{+0.6}$ & $58_{-15}^{+14}$ \\ 
         \hline
    \end{tabular}
    \label{tab:glitch_solutions}
        \tablefoot{The posteriors samples for these glitches can be found in a GitHub repository\footnote{\url{https://github.com/ezubieta/PSR-J0537-6910-recoveries}}. Uncertainties define $68\%$ of the most likely values in the posteriors.
    Glitch numbers follow the counting of RXTE glitches \citep{2018MNRAS.473.1644A}. Subtract 45 to recover the NICER count.}
\end{table*}

\section{Results}\label{sec: results}

\begin{figure*}%[h!]
    \centering
    \includegraphics[width=0.44\linewidth]{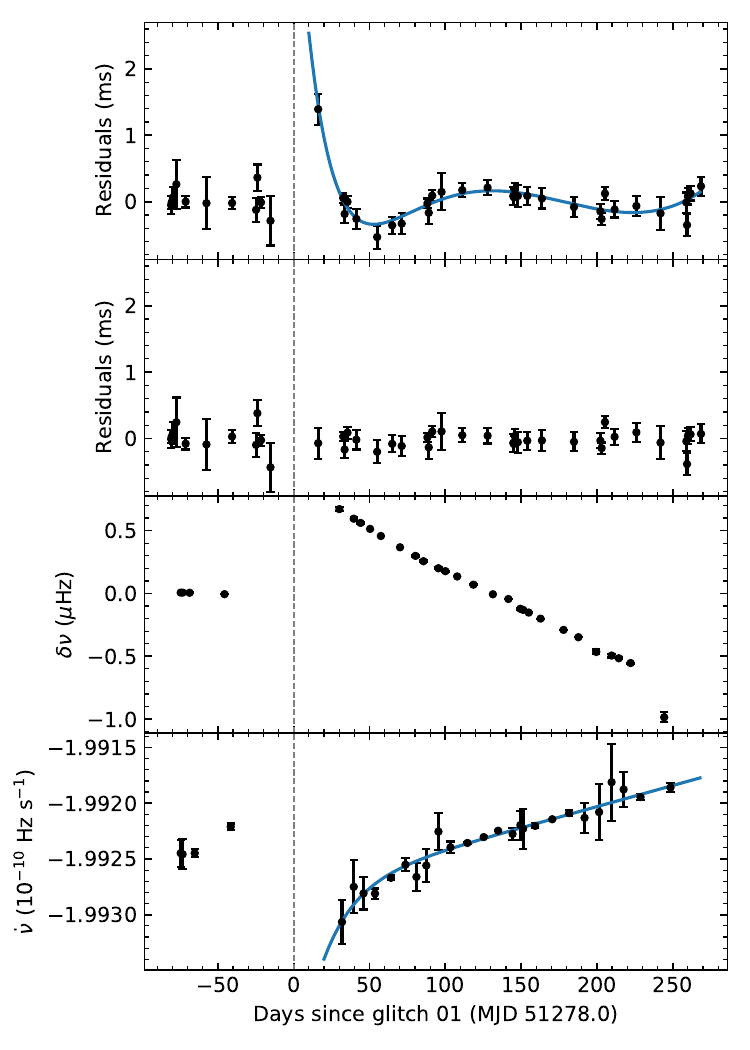}
    \includegraphics[width=0.44\linewidth]{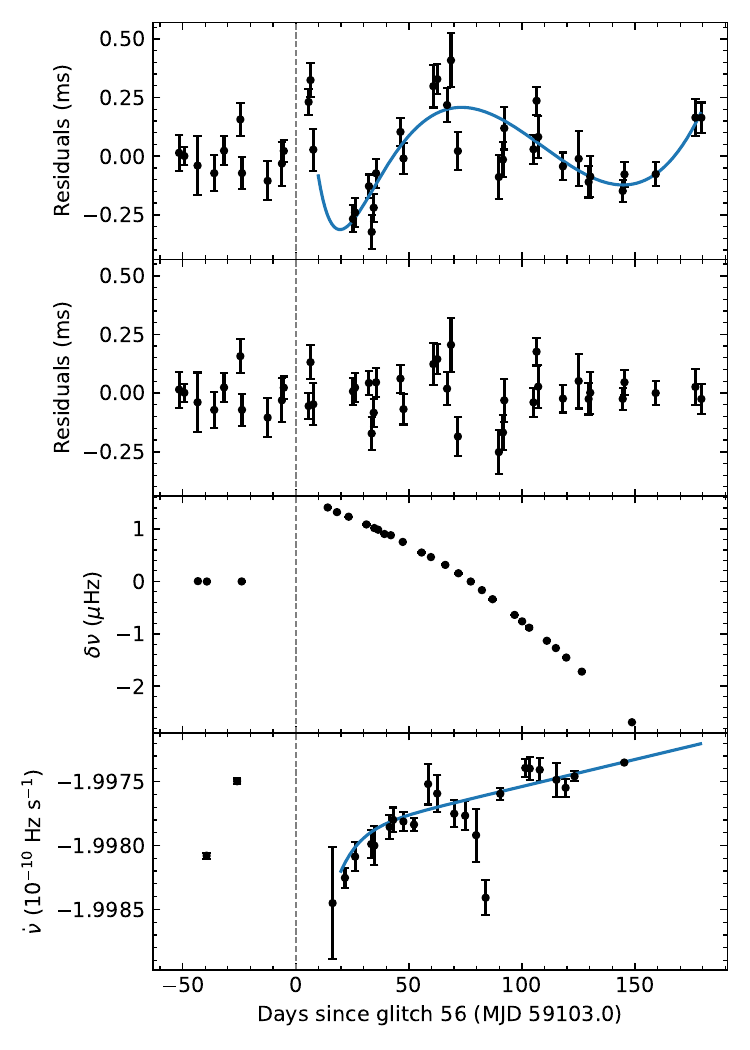}
    \caption{Examples of exponential glitch recovery detections, for glitches 1 (left) and 56 (right).
    The panels show, from top to bottom: Phase residuals relative to Model 1, with the residuals of the most likely exponential recovery model (Model 2 relative to Model 1) superimposed (blue curve); Phase residuals relative to Model 2; Frequency residuals relative to Eq.~\ref{eq:timing-model} fitted to TOAs up to the glitch epoch, with the post-glitch data all lowered by a certain amount (the mean post-glitch frequency residual) for better visualisation; $\dot\nu$ evolution with Model 2 shown as the blue curve.}
    \label{fig:detections}
\end{figure*} 

\subsection{Detection of six new glitches with NICER}

In the timing residuals of the most recent NICER data used in this analysis (MJD 59626 to 60601), the presence of six new glitches is evident -- and in accordance with the known glitching rate of this pulsar. 
We present the parameters that describe these new glitches in Table~\ref{tab:new_solutions}.

To be consistent with earlier measuring techniques and to ensure the results are directly comparable to previously published glitch parameters, we followed \citet{2020MNRAS.498.4605H} and used {\sc tempo2} \citep{2006MNRAS.369..655H} to fit Eq.~\ref{eq:timing-model}~and~\ref{eq:glitch-model} without the exponential relaxation term to the TOAs of each pre-glitch and post-glitch interval (described in Table~\ref{tab:new_solutions_description}). A seventh new glitch (number 67) occurred around MJD 60592, so we include glitch 66 in the analysis but not glitch 67, since the post-glitch data were incomplete at the time of this analysis (see \citealt{hoetal26}).

Five out of the six new glitches have inferred amplitudes $\Delta\nu_\mathrm{p}>15\,\mu$Hz and all had detectable changes in $\dot{\nu}$. As we discuss next, two of these glitches present evidence for exponential relaxation in their post-glitch recovery. Refined solutions that include this term are presented in the following subsection.

\subsection{Detection of exponential post-glitch relaxations}

The total of 66 glitches of \psr were analyzed following the methods outlined in Sect.~\ref{sec:methodology}, starting from initial solutions presented in \cite{2018MNRAS.473.1644A, 2020MNRAS.498.4605H, 2021ApJ...913L..27A,2022ApJ...939....7H} and Table~\ref{tab:new_solutions} of this work.
We found 12 glitches for which the Bayesian criterion points to the presence of an exponential recovery, with moderate evidence in two cases and strong or decisive evidence in the other ten. These glitches are presented in Table~\ref{tab:positive_detections}, together with information on the chosen solution for Models 1 and 2, and their Bayesian evidences. 

Some glitches yielded a clear $\tau_\mathrm{d}$ value during the initial systematic search, but their Bayesian evidence was not high enough to grant selection of Model 2 over Model 1.
This is likely due to the sparsity of the post-glitch TOAs in these particular cases.  
Nonetheless, in these cases where Model 2 was favoured, the accuracy of the inferred exponential solutions does not appear strongly sensitive to variations in data quality: there is no clear correlation between the uncertainties of the inferred exponential parameters ($\Delta\nu_\mathrm{d}$ and $\tau_\mathrm{d}$) and the TOA candence or TOA errors of the respective post-glitch interval.

As noticed in Table~\ref{tab:positive_detections}, in some glitches Model 1 required a $\Delta \ddot\nu_\textrm{p}\neq0$ in order to produce reasonably flat residuals, but when the recovery term (Model 2) was added, the change $\Delta \Ddot\nu_\mathrm{p}$ became negligible. There were also two cases (glitches 43 and 45), for which including a $\Delta \Ddot\nu_\mathrm{p}$ term did not significantly flatten the residuals for Model 1, but the exponential recovery term in Model 2 did. 
Finally, there were four cases for which both $\Delta \Ddot\nu_\mathrm{p}$ and a recovery term were required to significantly reduce the RMS of the residuals.
In the model of the first glitch, we omitted $\Delta \Ddot\nu_\mathrm{p}$, as there are only a few pre-glitch TOAs and therefore $\Ddot\nu$ cannot be reliably determined before the glitch. Similarly, $\Delta \ddot\nu_\textrm{p}$ was also not included for glitch 47, as only 3 pre-glitch TOAs were available.
In this case, $\Ddot\nu$ was fitted to both pre- and post- glitch data. 

The final solutions of Model 2 for the 12 glitches with evidence of an exponential term in their post-glitch recovery are shown in Table~\ref{tab:glitch_solutions}. The description of the observations used for each fit is presented in Table~\ref{tab:glitch_solutions_description}. It is worth mentioning that for glitches 12 and 47, a fraction of the posterior samples of $\Delta\nu_\mathrm{d}$ is consistent with $0$, as indicated by the respective uncertainties. 
Nevertheless, we consider them significant detections because the modes of the posteriors are far from zero, and the Bayesian criterium favours an exponential term. 
These two glitches have short inferred $\tau_\mathrm{d}$ values, and so the lack of TOAs immediately after the glitch hampers their accurate measurement, as reflected by their uncertainties. The shorter ($\sim 3\mathrm{d}$) recovery timescales seen in Table~\ref{tab:glitch_solutions} emphasize the need to monitor \psr with high resolution immediately after its glitches, to detect and resolve such fast transients. Such observations will also improve our detection capability for exponentials of any timescale.

\subsection{The exponential post-glitch relaxations}

In Fig.~\ref{fig:detections}, we show two of the clearest detections of exponential relaxations in \psr. For completeness, in Fig.~\ref{fig:no-detections} we also show two large glitches for which we found no evidence of exponential signatures. %The full list of 
Plots for all 12 glitches are available in a GitHub repository\footnote{\url{https://github.com/ezubieta/PSR-J0537-6910-recoveries}}.

The exponentially-relaxing component is only a small fraction of the instantaneous change in spin frequency at the glitch. 
The inferred decaying amplitudes $\Delta\nu_\mathrm{d}$ range between 0.05~and~0.8~$\mu\mathrm{Hz}$, while the persisting changes $\Delta\nu_\mathrm{p}$ lie between 20~and~43~$\mu\mathrm{Hz}$. Therefore, the percentage of the initial frequency step that decayed exponentially $Q$ is $<4\%$ for all glitches. Most often, the calculated $Q$ is close to or below $1\%$, which is typical of giant glitches in other pulsars  \citep[see, for instance,][]{2023MNRAS.521.4504Z}.

It is remarkable that among the glitches with an exponential recovery, two (no. 43 and 45) did not exhibit persistent changes $\Delta\dot\nu_\mathrm{p}$ nor $\Delta\Ddot\nu_\mathrm{p}$. For these glitches, the value of $Q'$ reaches $100\%$, meaning $\dot\nu$ returned to its pre-glitch value once the exponential decay was completed. 
Similar high $Q'$ values are obtained for glitches 12 and 38. This is somewhat unusual for Vela-like glitches, but has been observed in other pulsars, like the Crab.

\begin{figure}
    \includegraphics[width=\linewidth]{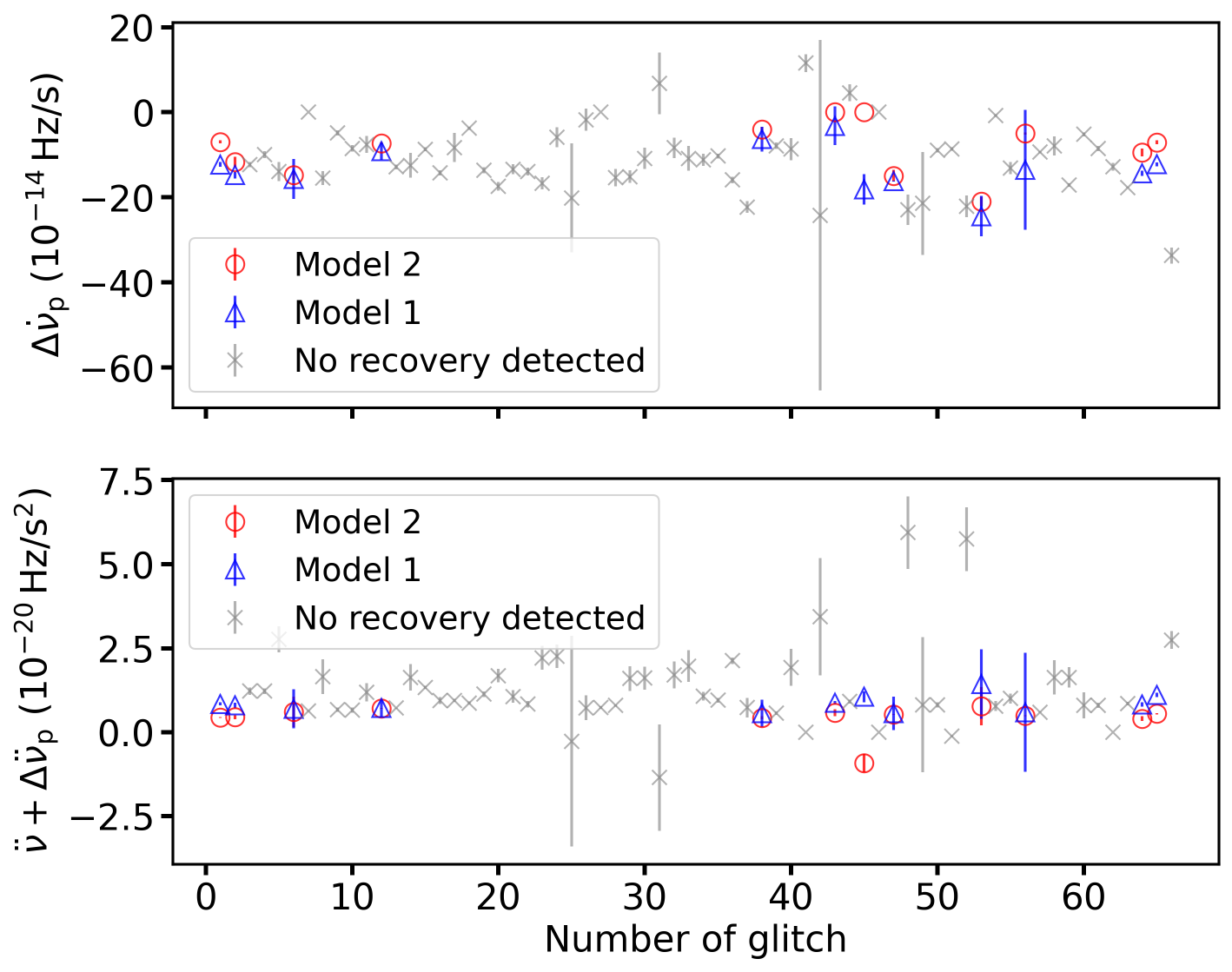}
    \caption{Difference in inferred glitch parameters between Model 1 (blue points) and Model 2 (red points), for glitches with detected exponential recovery.  Gray points represent values for glitches without detectable exponential terms (Model 1).}
    \label{fig: comparison}
\end{figure}

Although the persistent $\Delta\nu_\mathrm{p}$ does not change significantly with the addition of the exponential recovery term, the overall glitch and timing solution is altered. To investigate how including the exponentially recovering terms affects the inferred persisting  $\Delta\dot\nu_\mathrm{p}$ and $\Ddot\nu$, we compare their values as calculated by Model 1 (no exponential) and Model 2 (one exponential term) in Fig.~\ref{fig: comparison}. 
As can be seen, the persistent spin-down change $\Delta\dot\nu_\mathrm{p}$ considerably decreases when exponential relaxation is included in the model.  
Particularly interesting are the cases of glitch 43 and 45, discussed above. For these, Model 1 required a $\Delta\Dot\nu_\mathrm{p}\neq0$ term, which becomes zero for Model 2. This shows that, in some cases, exponential signatures may be misinterpreted as changes $\Delta\dot{\nu}_\mathrm{p}$ and $\Delta\ddot{\nu}_\mathrm{p}$.

\begin{figure}
    \includegraphics[width=\linewidth]{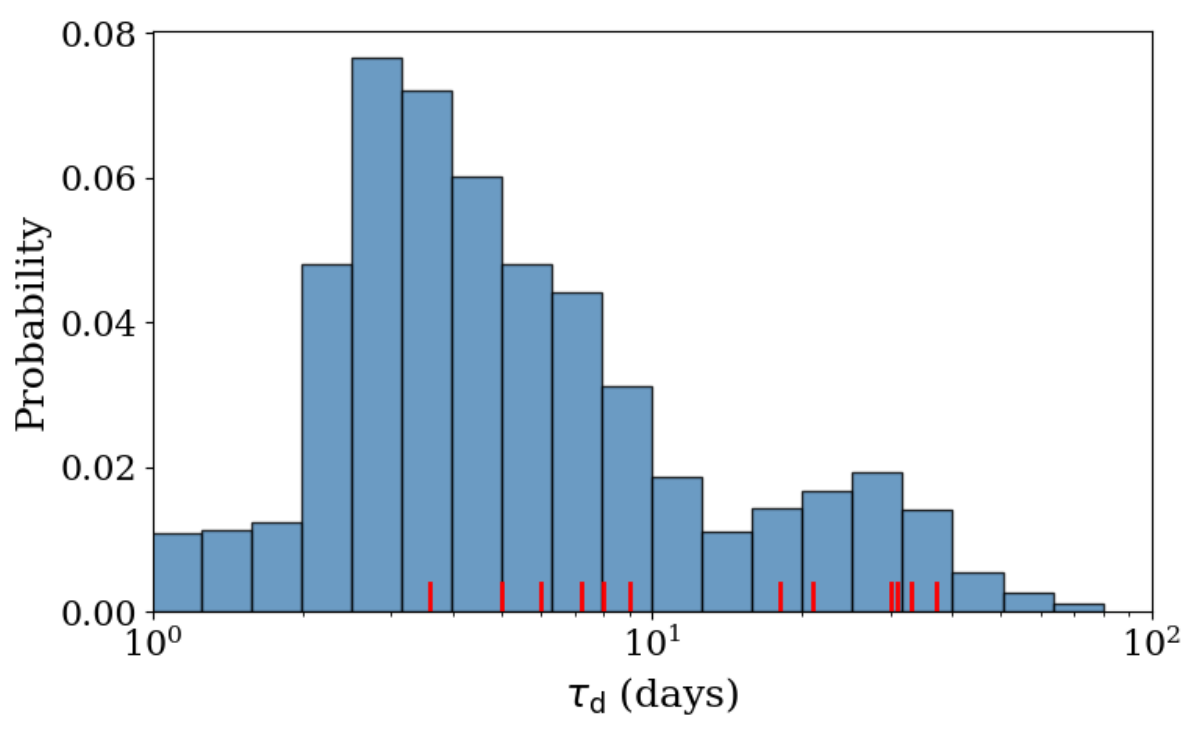}
    \caption{Distribution of $\tau_\mathrm{d}$ values for the 12 glitches with a detected recovery as represented by a histogram of 10000 samples (weighted) from the posterior distributions. The red ticks correspond to the values yield by the final solutions, as reported in Table~\ref{tab:glitch_solutions}.}
    \label{fig: tau_samples}
\end{figure}

The nominal timescales of the retrieved solutions for Model 2 span from 4~to~37~d, but are often highly uncertain (as seen in Table~\ref{tab:glitch_solutions}). To better capture the inferred range of typical recovery timescales of \psr, in Fig.~\ref{fig: tau_samples} we show the distribution of $\tau_\mathrm{d}$ values drawn from the posteriors of glitches with detected exponential recovery. There is some bimodality in the likely timescales, with a high peak around $\tau_\mathrm{d} \sim 3~\mathrm{d}$ and a secondary one around $\tau_\mathrm{d} \sim 20~\mathrm{d}$. We note here that for those glitches where the inferred timescale was relatively long (glitches 1, 2, 43, 45, 64 and 65), we also test a recovery model with two exponential terms but found no evidence for a second exponential. These results, which in general suggest short decaying timescales, intensify the need for dense observations as soon as possible after large glitches of \psr, in order to capture their short-term recovering behaviour. 

Higher observing cadence close to a glitch will also allow us to place better limits at its size and improve the overall accuracy of all glitch parameters and timing solutions. Despite the small amplitude $\Delta\nu_{\rm d}$ of the exponential relaxations compared to the total glitch change $\Delta\nu$, detecting and including these terms, when present, can affect several other results. 
This can be seen, for example, in Fig.~\ref{fig: time_next_glitch} that displays the size $\Delta\nu$  of a glitch versus the length of the time interval $\Delta T_{\rm f}$ until the following glitch.  
For \psr these parameters are known to exhibit a strong, rather linear (at least for intermediate to large glitches), correlation \citep{2006ApJ...652.1531M,2018MNRAS.473.1644A,2020MNRAS.498.4605H,2022ApJ...939....7H}. Here, we study the correlation between these two parameters through the Spearman rank-order correlation coefficient (\textsc{spearmanr} from the \textsc{scipy} library). We first consider all glitch amplitudes obtained via Model 1 (no exponential term, and find a correlation coefficient $\rho_s=0.89$ with a \textit{p}-value  $p=1.5\times10^{-22}$.  Interestingly, when we instead use the sizes $\Delta\nu$ retrieved by Model 2 for the 12 glitches with evidence of exponentials, the correlation measures improve to $\rho_s=0.94$ and \textit{p}-value of $p=2.0\times10^{-32}$.

\section{Discussion}\label{sec: discussion}

\begin{figure}%[h!]
    \includegraphics[width=\linewidth]{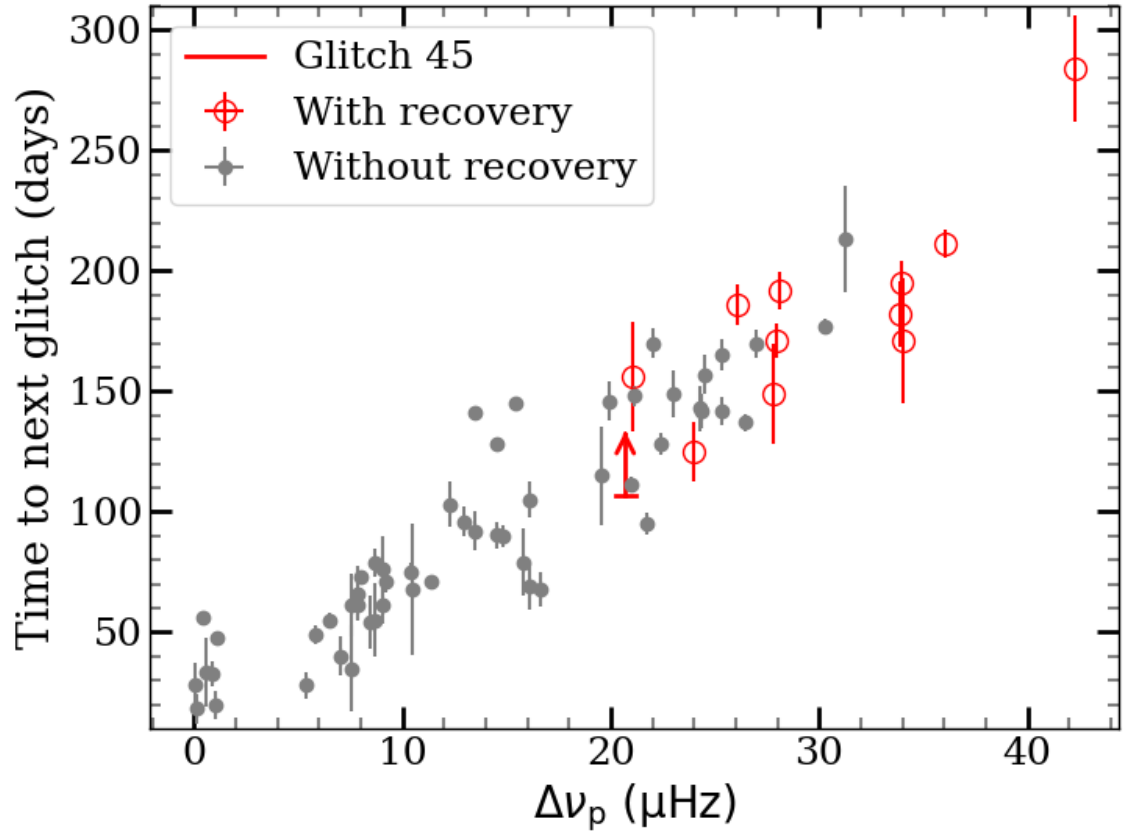}
    \caption{Time to the next glitch as a function of $\Delta \nu_p$, for 66 glitches of \psr. Red markers show glitches for which an exponential recovery was detected. Using the Model 2 solutions in these cases improves the size-waiting time correlation.
    Glitch 45, with a detected recovery, is the last glitch in the RXTE dataset, which ends $107$\,d after. 
    Thus the time to the next glitch is unknown and it is plotted with a vertical arrow indicating the lower limit.
    The latest glitch that we report in this work is glitch 66, which we are able to include in this plot as we know glitch 67 occurred on MJD 60592.
    }
    \label{fig: time_next_glitch}
\end{figure}

Exponential recoveries were identified and measured following 12 glitches of \psr, all of which rank among the largest glitches observed. 
No evidence of exponential recovery was found for glitches with  $\Delta\nu < 20~\mu\mathrm{Hz}$. For greater glitch sizes, some -- but not all-- events show an exponential component, and only the largest glitches ($\Delta\nu \gtrsim 31.3~\mu\mathrm{Hz}$) consistently exhibit detectable exponential terms.
This behaviour can be appreciated in Fig. \ref{fig: time_next_glitch}.
By visually inspecting the timing residuals relative to a model as in Eq.~\ref{eq:glitch-model}, but without the exponential term, we confirm that no more glitches of sizes between 20~and~30~$\mu$Hz show obvious signs of an exponential recovery; although not all residuals are adequately flat either.  
For instance, the timing residuals for glitches 66 and 28, which are the first and second largest events without a detected recovery (with $\Delta\nu=31.3\,\mu$Hz and $\Delta\nu=30.3\,\mu$Hz, respectively), are sufficiently flat, similarly to what is observed for glitch 9 (the fourth largest glitch without a recovery, with $\Delta\nu=26.4$\,$\mu$Hz). 
However, glitch 51 ($26.9$\,$\mu$Hz), which is the third largest glitch without an exponential detection, exhibits noisier residuals relative to the same model. 
Other glitches within this size range behave similarly to these largest events, two of which are shown in Fig.~\ref{fig:no-detections}. 
We could not find any obvious reason, such as observational biases or failure to retrieve equally accurate solutions, as to why exponentials were not detected after these large glitches. Furthermore, the presence of a detectable exponential relaxation seems associated with lower inter-glitch $\ddot{\nu}$ values, as discussed below. It is thus possible that the differences in post-glitch behaviour are intrinsic to the pulsar. Future observations can reveal whether that is the case and what it depends upon.

\subsection{Inter-glitch spin-down rate evolution}

\begin{figure}
    \includegraphics[width=\linewidth]{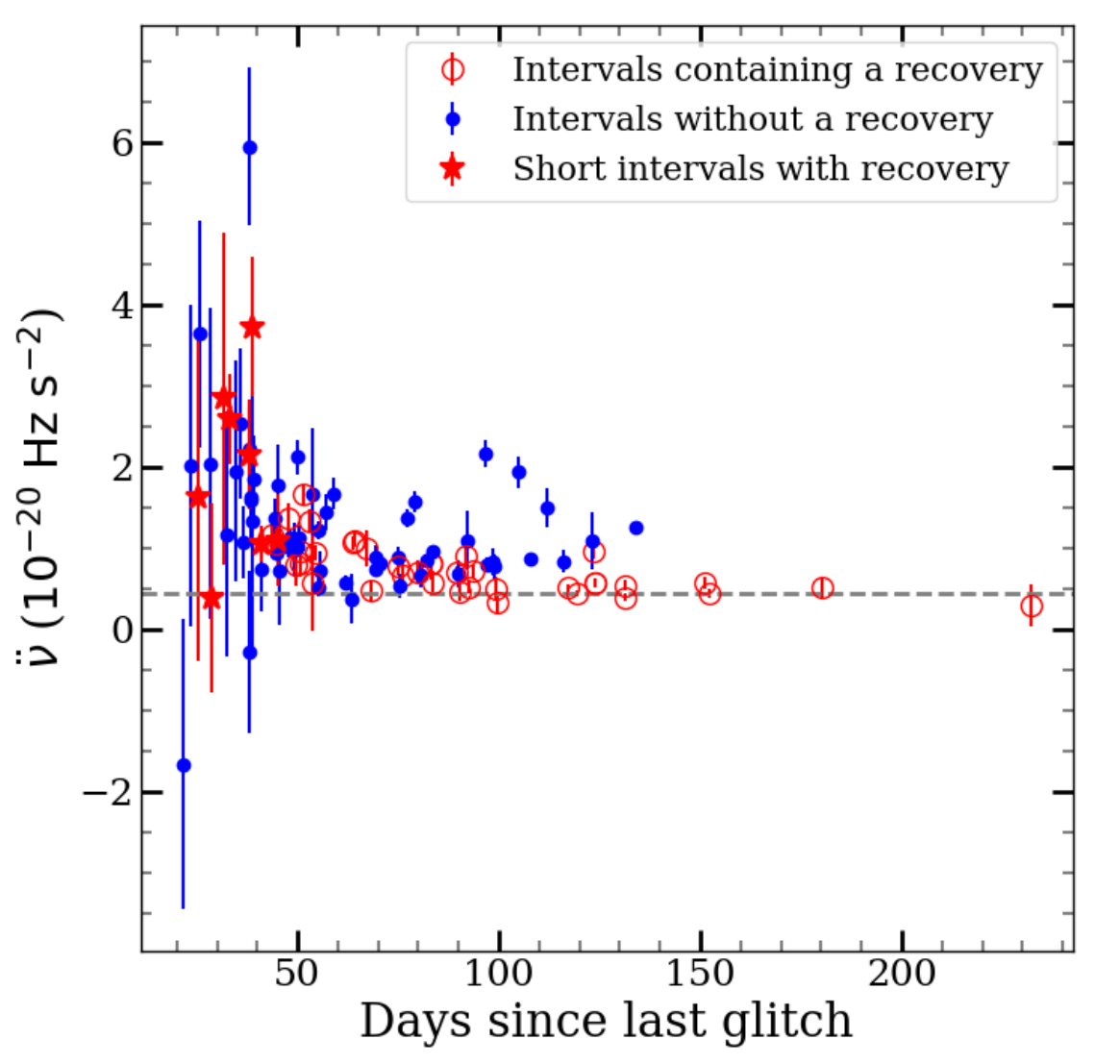}
    \caption{$\ddot\nu$ versus time since the preceding glitch. The dashed horizontal line indicates $\ddot\nu_\mathrm{ref} = 0.44\cdot10^{-20}~\mathrm{s^{-3}}$.
    The data points were obtained by fitting Eq. \ref{eq:timing-model} over a 90-day sliding window, which moved forward 20 days at each step. 
    The full inter-glitch data span was fitted if the inter-glitch interval was shorter than 90 days. 
    The red stars were obtained by fitting smaller windows of at least $30~\mathrm{d}$ and at least 8 TOAs right after the glitches with a recovery detected.}
    \label{fig: F2_plot}
\end{figure}

When the favoured glitch model includes an exponential term, the post-glitch rotation typically evolves with a lower $\ddot\nu$.
This is not surprising at first, as all detections concern large glitches: these are followed by long (at least $100$\,d) inter-glitch intervals due to the aforementioned $\Delta\nu$-$\Delta T_{\rm f}$ correlation, and it has been shown that the inter-glitch $\ddot\nu$ tends to decrease over time from the glitch epoch \citep{2018ApJ...864..137A}. To test this with our updated glitch sample, we fit Eq.~\ref{eq:timing-model} to 90-days TOA intervals, starting after each glitch and moving forward by 20 days until the next glitch is reached. The resulting $\Ddot\nu$ values as a function of time since the preceding glitch are seen in Fig.~\ref{fig: F2_plot}. 
When all glitches are considered collectively,  $\Ddot\nu(t)$ indeed appears to decrease over time. However, beyond $\sim70\,$d the glitches without a detectable exponential have higher $\Ddot\nu$ (blue points) compared to glitches with exponential relaxation (red circles).  
We emphasize that the modeled exponential contributions have not been removed before calculating these $\Ddot\nu(t)$ values. If that was the case, the red markers in Fig.~\ref{fig: F2_plot} (circles and stars) would be even lower, especially at early times. 
Therefore, we do not expect the blue points to be higher at late times due to unidentified exponentials of similar timescales as the detected recoveries. 
Although we cannot exclude contributions from  undetected exponential relaxations on much longer timescales, given the current analysis and the broad range of trial timescales this explanation seems considerably less likely than a different recovering regime. 

This genuine difference between glitches with and without detected exponential terms can also be expressed via the braking index. Assuming a power law $\dot{\nu}\propto-\nu^n$ for the spindown evolution, the braking index $n$ can be calculated from observations as $n=\nu \ddot{\nu}/\dot{\nu}^2$. 
Typical inter-glitch braking indices $n_\mathrm{ig}$ in Vela-like pulsars are $n_\mathrm{ig}=10-40$, or higher; arguably the result of internal superfluid torques \citep{2017MNRAS.466..147E,2018MNRAS.473.1644A} rather than external, magnetospheric torques which are expected to give much lower values (e.g. $n\sim3$ in the standard magnetic dipole braking model). Whilst for most glitches of \psr the interglitch $n_\mathrm{ig}$ is generally above 9 and as high as 35, after the glitches with detected exponential recoveries it settles to lower values, between 6.2 and 9.4.

To further investigate the $\ddot\nu$ values between glitches, and the overall rotational evolution, we study $\dot\nu$ as a function of time.
A $\dot\nu(t)$ dataset was constructed by fitting Eq.~\ref{eq:timing-model} to overlapping subsets of data that contain at least 5 TOAs and span at least $30$\,d, but not through glitches. For these fits we kept $\ddot\nu$ fixed at the mean value of each inter-glitch interval.  
The fitting window moved one TOA each time; hence, the data are heavily smoothed for visualization purposes.
In Appendix \ref{sec:appendix-dotnu} we present the resulting $\dot\nu$ evolution that resembles a sawtooth curve, typical of pulsars with repeating large glitches \citep[e.g.][]{2017MNRAS.466..147E}. In the case of \psr, $\dot\nu$ decreases over the years, with a corresponding effective negative slope and long-term negative effective braking index; a potential product of the high glitch activity \citep{2018MNRAS.473.1644A,2020MNRAS.498.4605H}. 
Between glitches, $\dot{\nu}$ increases rapidly: the interglitch evolution has high, positive $\ddot{\nu}$ values and $n_\mathrm{ig}>3$, 
which likely involves contributions from (at least) the exponentially and non-exponentially recovering superfluid, as well as the external electromagnetic and magnetospheric torques.  

\begin{figure}
    \includegraphics[width=\linewidth, trim=0.4cm 0.73cm 0.8cm 0cm, clip]{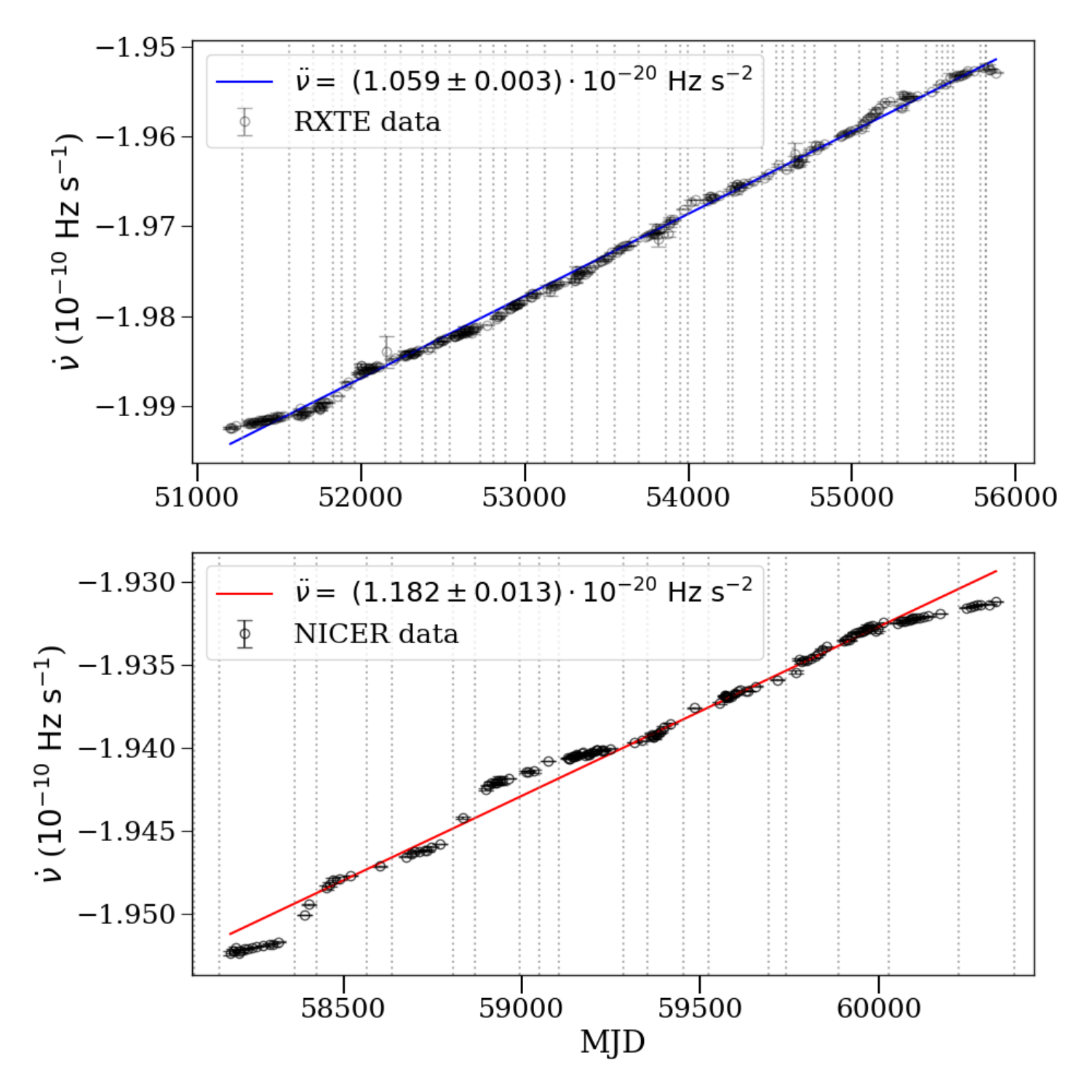}
    \caption{Evolution of $\dot\nu$ corrected for the changes induced by the glitches. Vertical dotted lines indicate the epoch of the glitches. Note that the scales on the horizontal axes are of different lengths.}
    \label{fig: F1-evolution}
\end{figure}

 The overall stability of the inter-glitch $\ddot{\nu}$ can be observed in Fig.~\ref{fig: F1-evolution}, where the modeled exponential and persisting glitch components are \emph{`corrected'} for, by
 subtracting $[\Delta\dot\nu_\mathrm{p} - \exp(-(t-t_g)/\tau_\mathrm{d})\Delta\nu_\mathrm{d}/\tau_\mathrm{d}]$ from $\dot{\nu}(t)$ after each glitch. 
This results in the disappearance of the saw-tooth curve and the long-term negative slope of $\dot{\nu}$ observed in Fig.  \ref{fig: spin-down}, leaving a predominantly linear, smooth evolution under a remarkably stable positive slope ($\ddot{\nu}>0$). This high interglitch $\ddot{\nu}$ and some additional wandering features (above random scatter) seen in Fig.~\ref{fig: F1-evolution} remain unaccounted for by the current glitch model which has only one exponential term to describe internal torques and does not, e.g., include the response of non-linearly coupled superfluid components \citep{2020MNRAS.499..161H}.    
We performed separate linear fits for the RXTE and NICER data and found $\Ddot\nu=(1.059\pm0.003)\cdot10^{-20} ~\mathrm{Hz}~\mathrm{s^{-2}}$ for RXTE and $\Ddot\nu=(1.182\pm0.013)\cdot10^{-20} ~\mathrm{Hz}~\mathrm{s^{-2}}$ for NICER. The  corresponding braking indices are high ($n_\mathrm{ig}=16.49\pm0.05$ and $n_\mathrm{ig}=18.4\pm0.2$, respectively) as in other similar glitching pulsars. 

\begin{figure}
    \includegraphics[width=\linewidth]{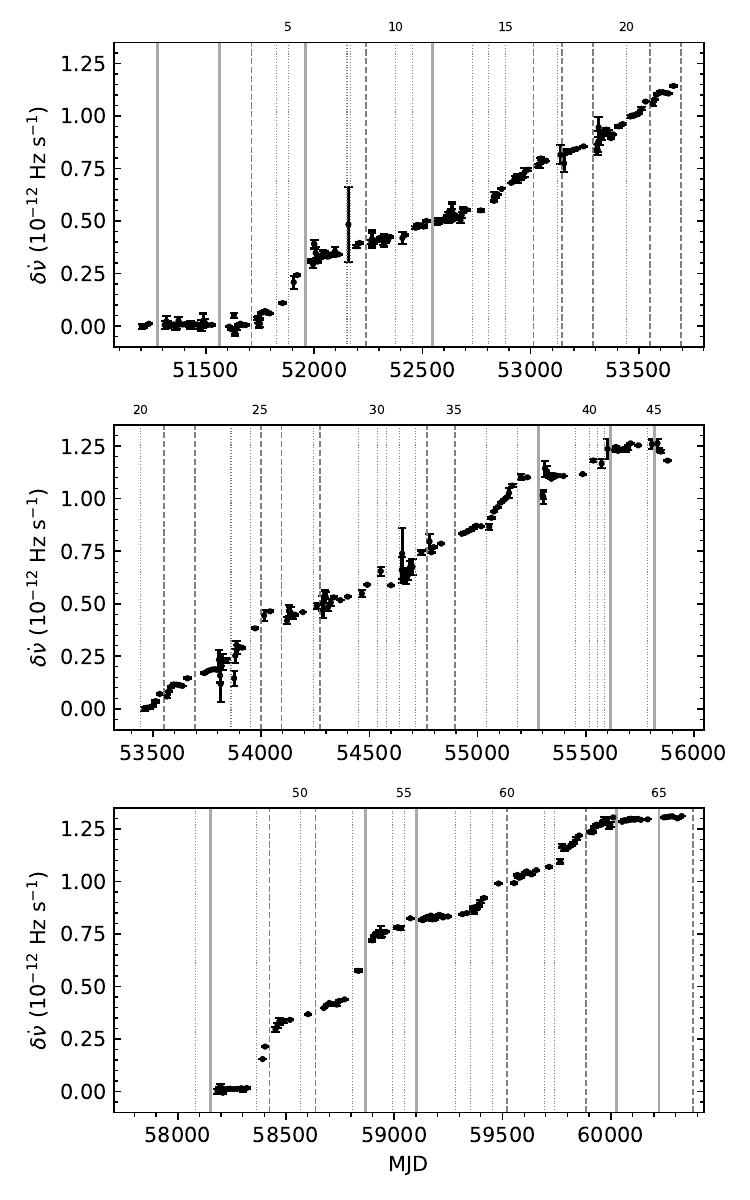}
    \caption{Residuals of the $\dot\nu$ evolution (as calculated for Fig. \ref{fig: F1-evolution}, with the modeled glitch persistent change and exponentially-recovering components removed), relative to a linear model with slope $\ddot\nu=0.44\times 10^{-20}\,$Hz\,s$^{-2}$. The full dataset was divided in three sections of similar length, shown in the three panels (RXTE data in the two superior panels and NICER at the bottom).
    All plots have equal vertical and horizontal scales. The solid gray lines mark the epochs of the glitches with a detected exponential recovery, dashed lines mark the epochs of glitches with $\Delta\nu>19\,\mu$Hz and dotted thin lines represent the epochs of smaller glitches.
    Glitch numbers are indicated on top of each panel.}
    \label{fig:F1-resids2}
\end{figure}

\begin{figure}[h!]
    \includegraphics[width=\linewidth]{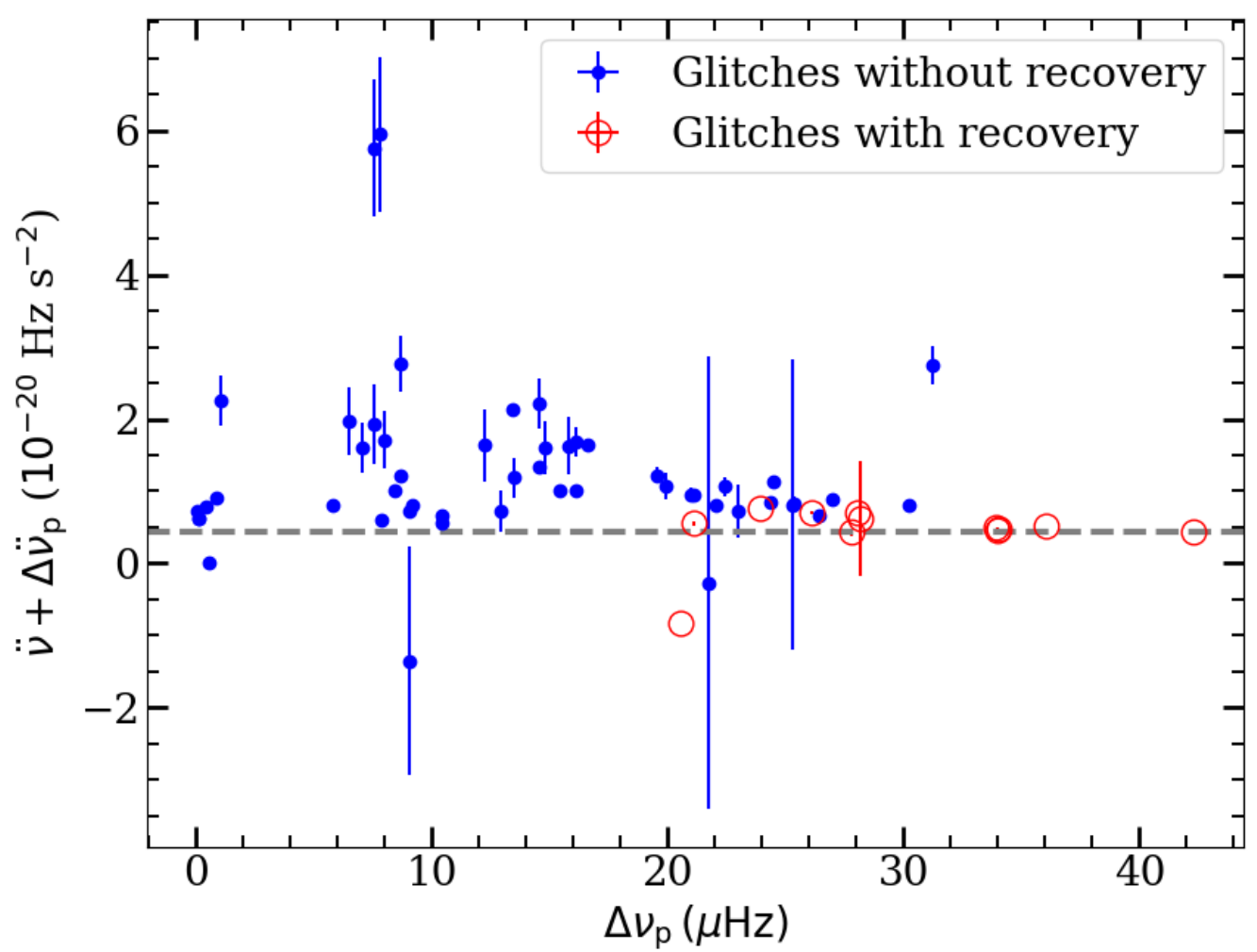}
    \caption{$\Ddot\nu+\Delta\Ddot\nu_\mathrm{p}$, as a function of $\Delta\nu_\mathrm{p}$ (that is, inter-glitch $\ddot\nu$ versus the size of the previous glitch) for all the 66 glitches. The dashed horizontal line indicates $\Ddot\nu=0.44\cdot10^{-20}~\mathrm{Hz}~\mathrm
    {s^{-2}}$.}
    \label{fig: F2-GLF0}
\end{figure}

To better understand the variations of $\ddot{\nu}$, we subtract (additionally to the corrections for the glitch components) a linear model with a fixed, reference value of $\ddot\nu_{\rm{ref}}=0.44\times 10^{-20}\,$Hz\,s$^{-2}$ which corresponds to one of the lowest inter-glitch measurements of $\ddot\nu$ (observed after the first and largest glitch; $n_\mathrm{ig}=6.8$).
These new residuals are shown in Fig. \ref{fig:F1-resids2}.
As expected, the $\dot{\nu}$ residuals are flat in the first interglitch interval, but they also remain relatively flat after the second glitch (large, followed by an exponential relaxation), indicating that $\ddot\nu$ stays close to $\ddot\nu_{\rm{ref}}$. However, around the time of the third glitch, which is large but has no detected exponential, $\ddot\nu$ increases and stays high until the next exponentially-recovering glitch occurs (glitch 6), after which it decreases again and stays a little above $\ddot\nu_{\rm{ref}}$ throughout several glitches. Similar patterns can be seen in the rest of the data: $\ddot{\nu}$ falls close to $\ddot\nu_{\rm{ref}}$ after glitches with detected exponentials, but overall spends long periods around a higher, rather constant, value. This is most clear in the first half of the middle panel, when many glitches occurred but none had detectable exponentials: $\ddot{\nu}$  stays high and relatively constant for about 2000 days.   Remarkably, some of the sharper $\ddot{\nu}$ increases appear in Fig. \ref{fig:F1-resids2} to occur around small glitches instead of intermediate/large ones (and never after an exponentially-recovering glitch).   
To explore this possible relation between the inter-glitch $\ddot\nu$ and the size of the previous glitch, we plot the value of $\Ddot\nu$ after each glitch, i.e. $\Ddot\nu+\Delta\Ddot\nu_\mathrm{p}$, against $\Delta\nu_\mathrm{p}$ of the preceding glitch. 
The result is shown in Fig. \ref{fig: F2-GLF0}, which shows a weak anti-correlation between $\Ddot\nu+\Delta\Ddot\nu_\mathrm{p}$ and $\Delta\nu_\mathrm{p}$ with the Spearman-rank correlation coefficient $\rho_s=-0.314$ and a $p$-value of $p=0.012$.

\subsection{$\Delta\dot\nu_\mathrm{p}$ and spin-down evolution}

\begin{figure}[h!]
    \includegraphics[width=\linewidth]{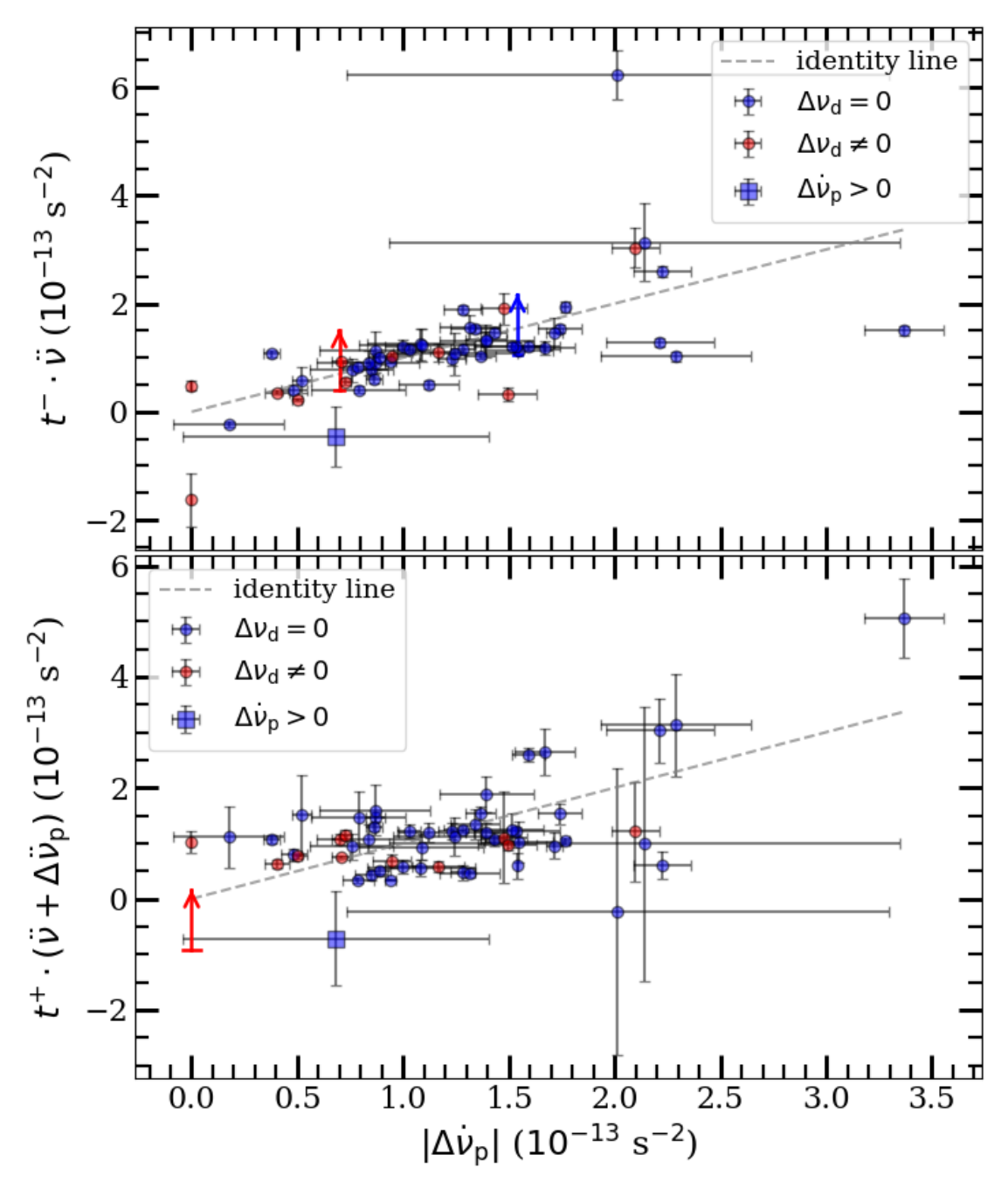}
    \caption{Comparison between $\Delta\dot\nu_\mathrm{p}$ and the waiting time ($t^{\pm}$) multiplied by value of the second derivative of the frequency ($\Ddot\nu$ before the glitch, $\Ddot\nu+\Delta\ddot\nu_\mathrm{p}$ after the glitch). Red points represent the glitches for which recoveries were detected, and blue points glitches for which we did not detect a recovery. The blue square represents a glitch for which we measured a positive value of $\Delta\dot\nu_\mathrm{p}$. Top panel:  $t^-$ is the time since the previous glitch. The vertical arrows show the lower limits for glitches \#1 (red arrow) and \#46 (blue arrow). Bottom panel: $t^+$ is the waiting time until the following glitch. The vertical arrow shows the lower limit for glitch \#45. As in Fig. \ref{fig: time_next_glitch}, we included glitch \#66 in this plot.}
    \label{fig: corr}
\end{figure}

As the frequent, large persisting changes of the spin-down rate $\Delta\dot\nu_\mathrm{p}$ are the most probable cause of the observed negative long-term trend of the spin-down evolution (Fig. \ref{fig: spin-down}), we explore their correlations with the change of $\dot{\nu}$ during the time intervals preceding and following the glitch. The results can be seen in Fig. \ref{fig: corr}. 
We calculate the change of spin-down rate during the inter-glitch intervals as $\Ddot\nu \cdot \mathrm{t}^-$ for the pre-glitch data span and $(\ddot\nu +\Delta\Ddot\nu_\mathrm{p}) \cdot \mathrm{t}^+$ for the post-glitch data span. Here, $\mathrm{t}^-$ refers to the time since the previous glitch and $\mathrm{t}^+$ refers to the time to the next glitch. Following \citet{2024MNRAS.532..859L}, the smallest glitches ($\Delta\nu/\nu<10^{-7}$) were excluded from the analysis and the waiting times were recalculated between the remaining glitches. We find a significant correlation between the increase of $\dot{\nu}$ during an interglitch interval and its change $|\Delta\dot{\nu}_\mathrm{p}|$ at the subsequent glitch, with $\rho_s=0.79$ and $p$-$\mathrm{value}=9.3\cdot10^{-13}$. This is in accordance with previous results for this particular pulsar \citep{2006ApJ...652.1531M,2018MNRAS.473.1644A}. On the other hand, a correlation of $|\Delta\dot{\nu}_\mathrm{p}|$  with the change of $\dot{\nu}$ in the time until the next glitch is not as tight, with $\rho_s=0.29$ and $p$-$\mathrm{value}=0.03$, although such a correlation has been observed for several other pulsars \citep{Lower:2021rdo,2024MNRAS.532..859L}. It was already noted in \cite{2024MNRAS.532..859L} that both correlations are seen for some glitching pulsars, and that for \psr $|\Delta\dot{\nu}|$ is more strongly correlated to the preceding (rather than following) increase of $\dot{\nu}$. Our results, which use the persisting changes (instead of the total change, which includes the exponentially relaxing components) further strengthen this conclusion. Because $\Ddot\nu \cdot \mathrm{t}^-$ is a proxy of the amount of superfluid that re-coupled since the previous glitch, and therefore is available to decouple again at the next glitch causing a bigger $\Delta\dot\nu$, this correlation implies a `reservoir' effect for the glitch $\dot{\nu}$ changes in \psr  \citep{2018MNRAS.473.1644A,2024MNRAS.532..859L}.

\subsection{Distributions of $\Delta\nu_\mathrm{p}$, $\Delta\dot\nu_\mathrm{p}$ and waiting times}

The number of glitches studied in this work (66) might be small by statistical standards, but it is the largest glitch sample from a single pulsar ever investigated. We therefore explore the distributions of basic glitch parameters ($\Delta{\nu}_\mathrm{p}$, $\Delta{\dot{\nu}}_\mathrm{p}$, and interglitch waiting times) using our updated measurements. The unbinned data were modeled using the \textsc{stats} package of the \textsc{scipy} library. For visualization purposes, we bin the data using the minimum histogram bin width between the Sturges rule and the Freedman-Diaconis estimator. 

\begin{figure}[h!]

\includegraphics[width=0.9\linewidth]{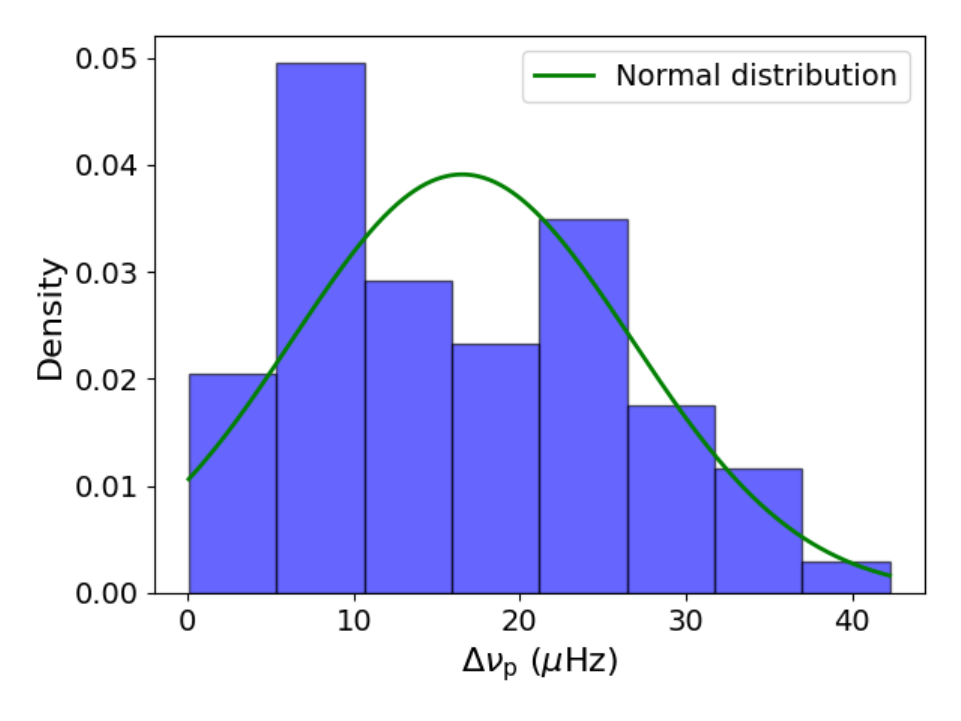}\\\includegraphics[width=0.9\linewidth]{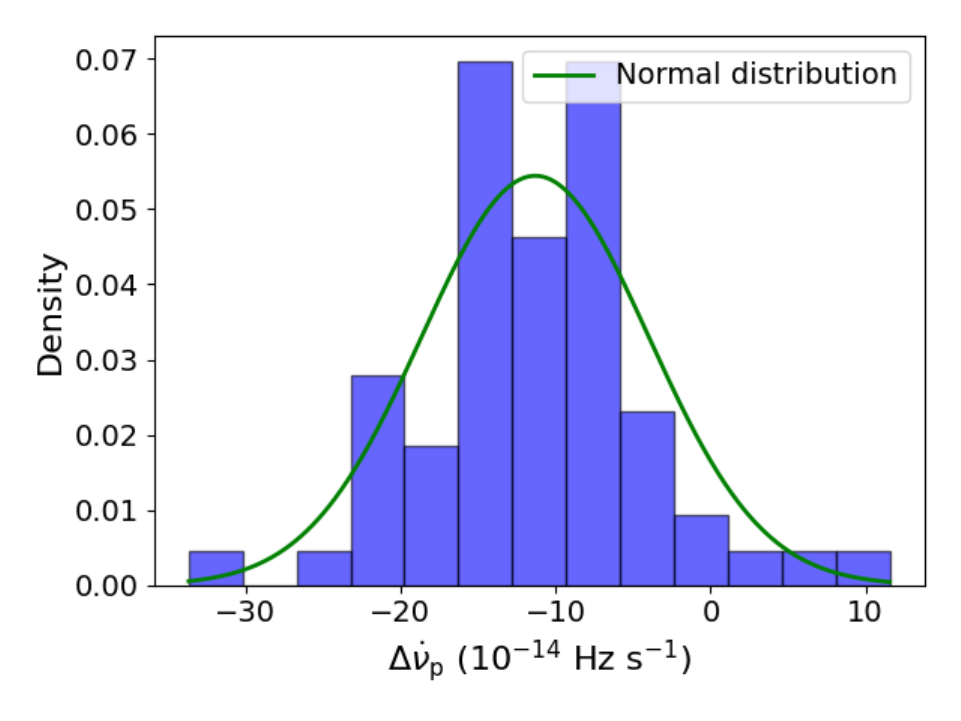}\\\includegraphics[width=0.9\linewidth]{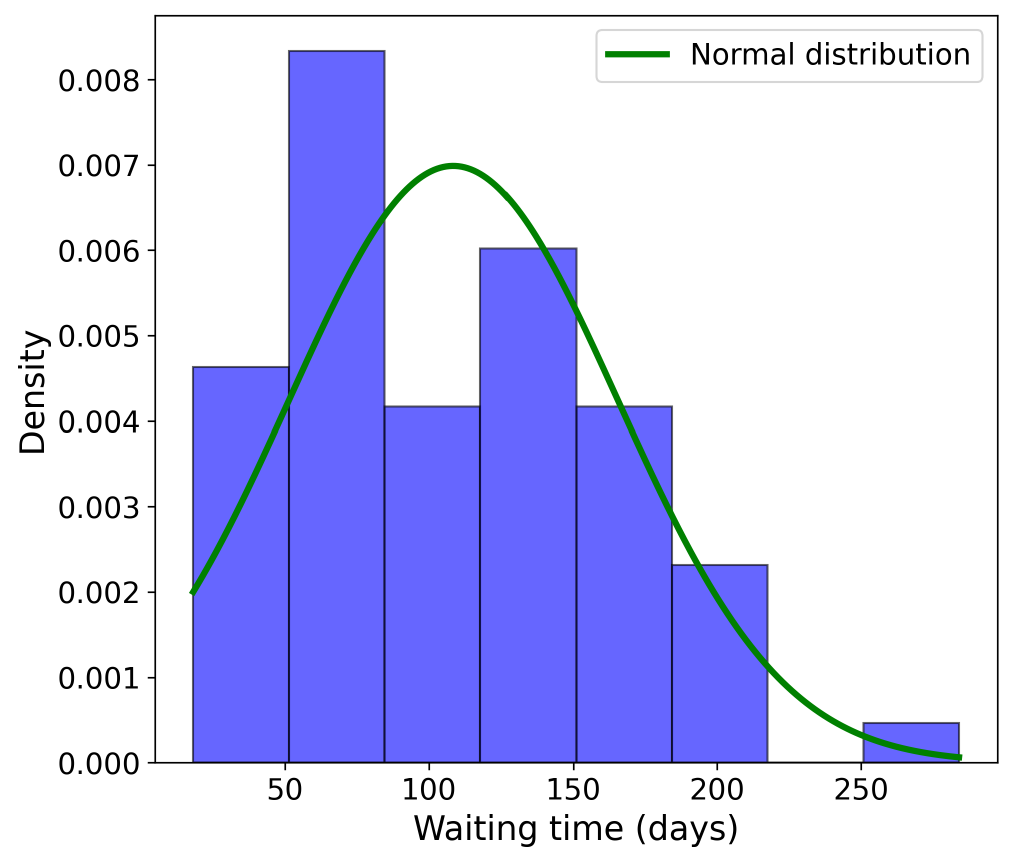}    
    \caption{Distribution of $\Delta\nu_\mathrm{p}$ (top panel), $\Delta\dot\nu_\mathrm{p}$ (middle panel) and the waiting times (bottom panel) of the glitches studied in this work. $\Delta\nu_\mathrm{p}$ and $\Delta\dot\nu_\mathrm{p}$ are fitted with a normal distribution; the waiting times with a normal and a bimodal distribution.}
    \label{fig: dist_GLF0}
\end{figure}

Histograms of the glitch parameters are shown in Fig.~\ref{fig: dist_GLF0}, together with the fitted models for Gaussian distributions, which are an acceptable fit to the data.

\begin{table}[!h]
    \setlength{\tabcolsep}{4pt}
        \caption{Description of the normal distributions of $\Delta\nu$, $\Delta\dot\nu$ and the waiting times.}
    \begin{tabular}{cccc}

        \hline\hline
          & $\Delta\nu_{\mathrm{p}}$   & $\Delta\dot\nu_{\mathrm{p}}$  & Waiting time  \\

        \hline
        KS & 0.10 & 0.09 & 0.12 \\
        
        $p$-value & 0.48 & 0.54 &0.22  \\

        $\mu$ & $16(1)$ & $-11.3(9)$ & 108(7) \\

        $\sigma$ & $10.2(9)$ & $7.3(7)$ & 57(5)  \\

         \hline
    \end{tabular}
        \label{tab:distributions}
        \tablefoot{KS is the result of the Kolmogorov-Smirnov test, $\mu$ is the center of the distribution and $\sigma$ its width. For $\Delta\dot\nu_\mathrm{p}$, $\mu$ and $\sigma$ are in units of $10^{-14}~\mathrm{Hz}~\mathrm{s^{-1}}$. For $\Delta\nu_\mathrm{p}$, they are expressed in $\mu\mathrm{Hz}$, and, for the waiting time, they are expressed in days. For $\mu$ and $\sigma$, the numbers in parenthesis are their uncertainties in the last digit.}
    \end{table}

We find that $\Delta\nu_\mathrm{p}$ and $\Delta\dot\nu_\mathrm{p}$ are consistent with normal distributions centered at $16~\mu\mathrm{Hz}$ and $-11.3 \times10^{-14} ~\mathrm{Hz}~\mathrm{s^{-1}}$, respectively, with $p-\mathrm{value}$ of 0.48 and 0.54. 
With respect to the inter-glitch waiting times, the $p$-value of the normal distribution is 0.22, and it is centered at $\mu=108~\mathrm{days}$ -- in accordance to what is seen in Fig. \ref{fig: time_next_glitch}, which shows that a waiting time of $\approx108~\mathrm{d}$ corresponds to a $\Delta\nu_\mathrm{p}\approx 16~\mu\mathrm{Hz}$. The parameters of the fitted distributions are presented in Table~\ref{tab:distributions}.

\section{Conclusion}
\psr is the most frequently glitching pulsar, exhibiting an average of three glitches per year. 
Analysis of some recent NICER observations delivered the detection of 6 new glitches, with properties typical for this pulsar (Table \ref{tab:new_solutions}).
With these new detections, the total number of published glitches for this pulsar is now 66.
The vast majority of the glitches in \psr are large compared to the majority of glitches detected in other pulsars, with 70\% of them having $\Delta\nu>10\,\mu$Hz.  
In other young pulsars, large glitches like these are often followed by exponential-like recoveries. 
However, in the case of \psr, only one exponential recovery had been reported despite the detection of more than 50 large glitches. 
We performed a systematic search for exponential recoveries after all major glitches of \psr and defined detection criteria based on model comparison and Bayesian evidence. 
Using this method, 11 new exponential recoveries were identified and characterised.

The presence of detectable exponential recoveries seems to depend on glitch size.
Every glitch with size $\Delta\nu$ above $31.3~\mathrm{\mu Hz}$ exhibits an exponential recovery, but only some glitches of amplitudes  
$20~\mathrm{\mu Hz}<\Delta{\nu}< 31.3~\mathrm{\mu Hz}$ do, 
and no exponentials are detected for any glitch of size below $20~\mathrm{\mu s}$. \psr has a uniquely strong correlation between glitch size and the length of its following interglitch interval. This could introduce an observational bias as, for example, glitches with sizes below $10~\mathrm{\mu s}$ are followed by only a $\sim 50$\,d interval, which might be too short to securely detect the exponential recovery. It cannot, however, account of the lack of detections for glitches with $\Delta\nu\gtrsim 20~\mathrm{\mu s}$ which we believe may be an intrinsic behaviour of the pulsar. 

For the detected recoveries, the measured $\Delta\nu_\mathrm{d}$ values are between $0.05~\mu\mathrm{Hz}$ and $0.8~\mu\mathrm{Hz}$, while $\tau_\mathrm{d}$ ranges between $3~\mathrm{d}$ and $37~\mathrm{d}$, which evidences the need for a high-cadence monitoring of this pulsar to thoroughly characterize its glitches. All the values of $Q$ are below $4\%$, while $Q'$ is always between $46\%$ and $100\%$.
Including the exponential recovery in the timing model affects the inferred parameters, especially $\Delta\nu_\mathrm{p}$ and $\Ddot\nu$. We re-evaluate known correlations of this pulsar using the new solutions, and find a very strong correlation between $\Delta{\nu}_\mathrm{p}$ and the time until the following glitch, and a strong correlation between the change of $\dot{\nu}$ over a pre-glitch interval $t^{-}$, calculated as $(\ddot{\nu}\cdot t^{-})$, and the change $|\Delta{\dot\nu}_\mathrm{p}|$ at the following glitch. We also find a mild anti-correlation between $\Delta\nu_\mathrm{p}$ and its post-glitch value of $\Ddot\nu$. 
Importantly, it was demonstrated that the spin-down rate $\dot{\nu}$ evolves at a lower $\ddot\nu$ after glitches with exponential recovery (once the exponential relaxation is over) compared to the generally stable evolution with higher $\ddot\nu$ following glitches without detected exponential decay.

Our models do not include a red noise component. While timing noise can certainly influence some of the inferred recovery parameters, or even mimic an exponential relaxation, the combination of theoretical expectations, consistent trends across glitches, and the distinct residual structures strengthens our confidence that the majority of the detected exponential components represent a real post-glitch behaviour.

These findings can help understand the mechanism behind glitches and the superfluid re-coupling process that follows them, and shed light at the interplay of various internal and externals torques in act for this pulsar. Continuous, high-cadence, observations of glitching pulsars are essential to study neutron star dynamics, and \psr is an excellent target for such monitoring due to its frequent, large glitches and unique characteristics. As we showed in this work, there are still underexplored aspects of its rotational behaviour, and fast post-glitch transients which require good time coverage to achieve further progress.

\begin{acknowledgements}
We thank Zaven Arzoumanian and Keith Gendreau for scheduling extra observations with NICER, which allowed the detection of some of the post-glitch recoveries reported here.  
CME acknowledges support from ANID/FONDECYT, grant 1211964. DA acknowledges support from an EPSRC/STFC fellowship ( EP/T017325/1).
WCGH acknowledges support through grant 80NSSC22K1305 from NASA. FG is a CONICET researcher. FG acknowledges support from PIBAA 1275 and PIP 0113 (CONICET).
We used computing resources made available by ANID Chile SIA, grant SA77210112 (PI: Pinto).

This work is supported by NASA through the NICER mission and the Astrophysics Explorers Program and uses data and software provided by the High Energy Astrophysics Science Archive Research Center (HEASARC), which is a service of the Astrophysics Science Division at NASA/GSFC.

\end{acknowledgements}

\bibliographystyle{aa} % style aa.bst
\bibliography{biblio} % your references biblio.bib
-------------------------
\appendix
\onecolumn

\section{Data intervals used to measure new glitches and exponential recoveries}

Table \ref{tab:new_solutions_description} gives details on the fits of six new glitches (Table \ref{tab:glitch_solutions}).
Table \ref{tab:positive_detections} shows the Bayesian evidence of Models 1 and 2 for each glitch where model 2 was favored.
Table \ref{tab:glitch_solutions_description} presents details of the fits of glitch exponential recoveries to 12 glitches (Table \ref{tab:glitch_solutions}). 

\renewcommand{\arraystretch}{1.25}

\begin{table*}[h]
    \centering
    \caption{Data used to model six new glitches (Table 
    \ref{tab:new_solutions}).}
    \begin{tabular}{ccccccc}
        \hline\hline
        Glitch  & START  & FINISH  & Preglitch TOAs & Postglitch TOAs & Residual RMS & $\chi^2_\mathrm{red}$  \\ 
          &  (MJD) & (MJD) &  &  & $(\mu\mathrm{s})$ &   \\ 
        \hline
        61 & 59529.5 & 59735.4 & 27 &  12 & 138.7 &  4.4 \\ 
        62 & 59695.9 & 59882.6 & 12 &  22 & 142.5 &  4.8 \\ 
        63 & 59745.7 & 60026.9 & 22 &  26 & 148.5 &  6.0 \\ 
        64 & 59888.5 & 60221.3 & 26 & 21 & 275.1 & 20.4 \\ 
        65 & 60071.8 & 60357.4 & 16 & 15 &  124.4 & 4.1 \\ 
        66 & 60287.7 & 60601.4 & 10 & 7 & 104.5 & 4.1 \\ 
         \hline
    \end{tabular}
    \label{tab:new_solutions_description}
    \tablefoot{Glitches \#64 and \#65 present clear post-glitch recovery signatures. We removed the TOAs affected by these recoveries from the preglitch data span of glitches \#65 and \#66 to obtain an accurate preglitch solution before characterizing the respective glitches.}
\end{table*}

\renewcommand{\arraystretch}{1.25}
\begin{table*}[ht]%[htbp]
\centering
    \caption{Bayesian evidence for the models corresponding to the 12 positive detections.}
    \begin{tabular}{cccccccc}
        \hline\hline
        Glitch & $t_g$ (MJD)  & $\Delta\Ddot{\nu}_\mathrm{p}$ in Model 1 & $\Delta\Ddot{\nu}_\mathrm{p}$ in Model 2 & Evidence Model 1& Evidence Model 2 & $\mathrm{ln}~B_{21}$ & Evidence criteria \\ \hline
        1 & 51278(16) & No & No & 318.96 & 355.63 & 16 & Decisive \\ 
        2 & 51562(15) & Yes & No & 285.94 & 292.45 & 2.82 & Moderate \\ 
        6 & 51960(5) & Yes & Yes & 248.69 & 265.81 & 7.48 & Strong \\ 
        12 & 52545(6) & Yes & No & 340.81 & 352.68 & 5.16 & Strong \\ 
        38 & 55280(4) & Yes & No & 161.71 & 181.80 & 8.72 & Strong \\ 
        43 & 55615(4) & No & No & 141.72 & 164.92 & 10.07 & Decisive \\
        45 & 55819(2) & No & No & 118.74 & 127.78 & 3.92 & Moderate \\ 
        47 & 58152(11) & No & No & 174.70 & 186.76 & 5.23 & Strong \\ 
        53 & 58868(5) & Yes & Yes & 124.58 & 166.76 & 18.31 & Decisive \\ 
        56 & 59103(6) & Yes & No & 240.66 & 328.32 & 38.07 & Decisive \\ 
        64 & 60027(1) & Yes & Yes & 29.60 & 348.15 & 138.34 & Decisive \\ 
        65 & 60223(3) & Yes & No & 222.24 & 246.29 & 10.44 & Decisive \\ 
         \hline
    \end{tabular}
    \label{tab:positive_detections}
\end{table*}

\begin{table*}[!h]%[htbp]
    \centering
        \caption{Description of the data used to measure exponential glitch recoveries (Table \ref{tab:glitch_solutions}).}
    \begin{tabular}{cccccccc}
        \hline\hline
        Glitch  & START  & FINISH  & Preglitch TOAs & Postglitch TOAs & Residual RMS & $\chi^2_\mathrm{red}$ & Time to next glitch \\ 
          &  (MJD) & (MJD) &  &  & $(\mu\mathrm{s})$ & & ($\mathrm{days}$) \\ 
        \hline
        1 & 51197.1 &  51546.7 & 11 & 32 & 104.6 & 0.9 & 284  \\
        2 & 51423.0 & 51705.2 & 18 & 18 & 123.9 & 1.32 & 149 \\
        6 & 51886.9 & 52144.1 & 10 & 26 & 147.4 & 1.42  & 192 \\
        12 & 52460.0 & 52717.4 & 14 & 28 & 124.9 & 1.23 & 186 \\
        38 & 55185.4 & 55444.8 & 9 & 15 & 200.0 & 2.27 & 171 \\
        43 &  55589.3 & 55786.1 & 6 & 15 & 157.2  & 1.6 & 171 \\
        45 & 55794.7 &  55926.9 & 5 & 11 & 122.9 & 1.2 & $\geq107$ \\
        47 & 58108.1 & 58348.8 & 3 & 21 & 89.6 & 2.9 & 211 \\
        53 &  58810.0 & 58990.2 & 6 & 17 & 93.5 & 3.8 & 125 \\
        56 & 59051.4 &  59282.4 & 10 & 31 & 82.2 & 2.0 & 182 \\
        64 & 59888.5 & 60221.3 & 26 & 21 & 120.3 & 4.1 & 195 \\
        65 & 60071.8 & 60357.4 & 16 & 15 & 77.4 & 1.6 & 156 \\
         \hline
    \end{tabular}
        \label{tab:glitch_solutions_description}

    \end{table*}

\section{Examples of glitches without a detected recovery}\label{sec:appendix-no-detection}

We present two examples of large glitches for which we did not detect exponential recovery. In these cases, the algorithm described in Sect. \ref{sec:methodology} either did not identify a minimum of $\chi^2_\mathrm{red}$ within 1 day to 4 times the length of the data span, or, if it did, the Bayesian evidence for Model 2 was not significant compared to Model 1.

\begin{figure*}[h!]
    \centering
    \includegraphics[width=0.44\linewidth]{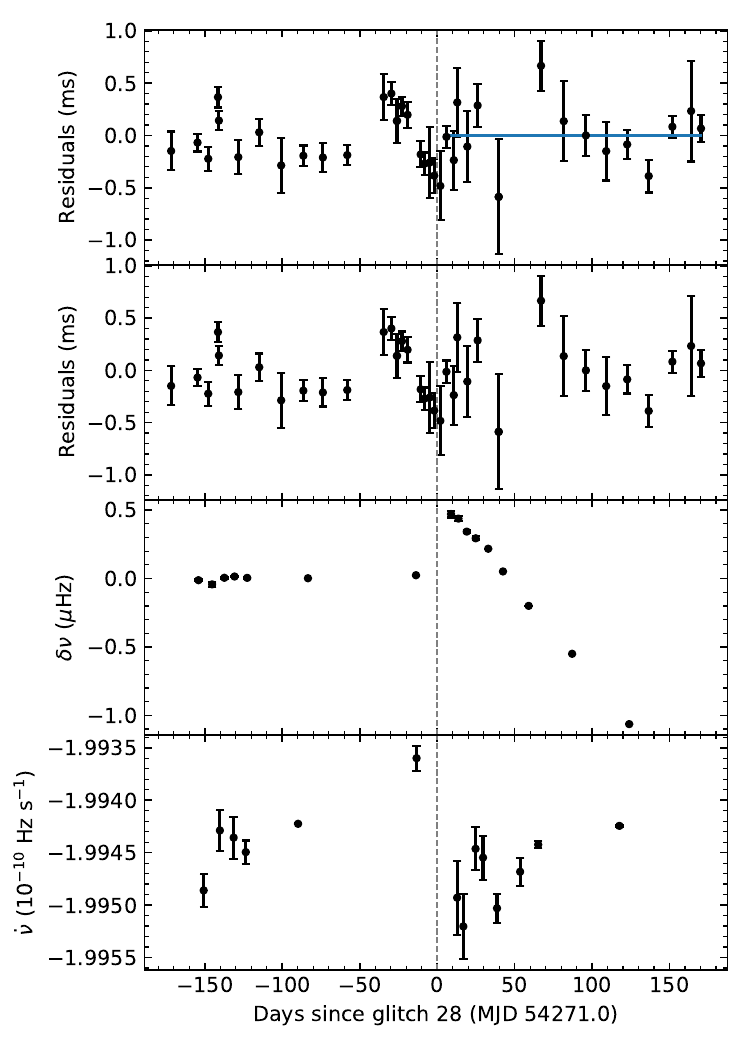}
    \includegraphics[width=0.44\linewidth]{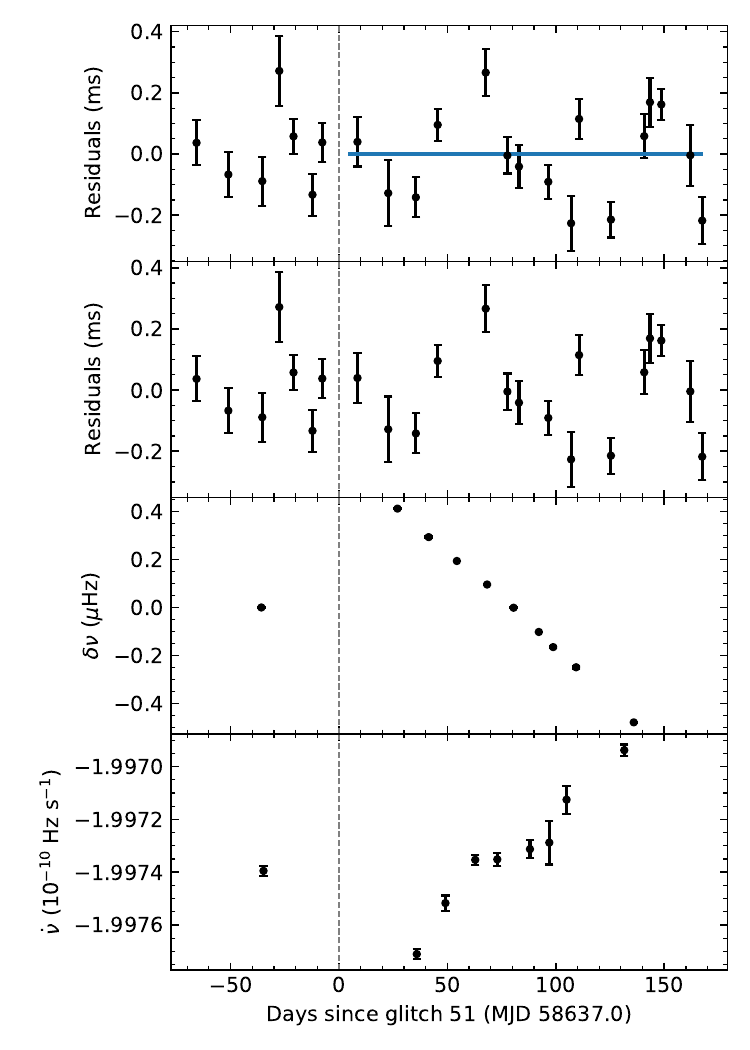}
    \caption{Examples of glitches 28 (left) and 51 (right) without 
    a detected exponential recovery.
    In each plot the panels show, from top to bottom: Phase residuals relative to Model 1, and superimposed (blue curve) the residuals of the most likely exponential recovery model (Model 2 relative to Model 1), which in this case corresponds to zero; Phase residuals relative to Model 2; Frequency residuals relative to Eq. \ref{eq:timing-model} fitted to TOAs up to the glitch epoch, with the post-glitch data all lowered by a certain amount (the mean post-glitch frequency residual) for better visualisation; $\dot\nu$ evolution with Model 2 shown as the blue curve when it was the favoured model.}
    \label{fig:no-detections}
\end{figure*} 

\clearpage

\section{Evolution of $\dot\nu$}\label{sec:appendix-dotnu}

\begin{figure}[h!]
    \includegraphics[width=\linewidth]{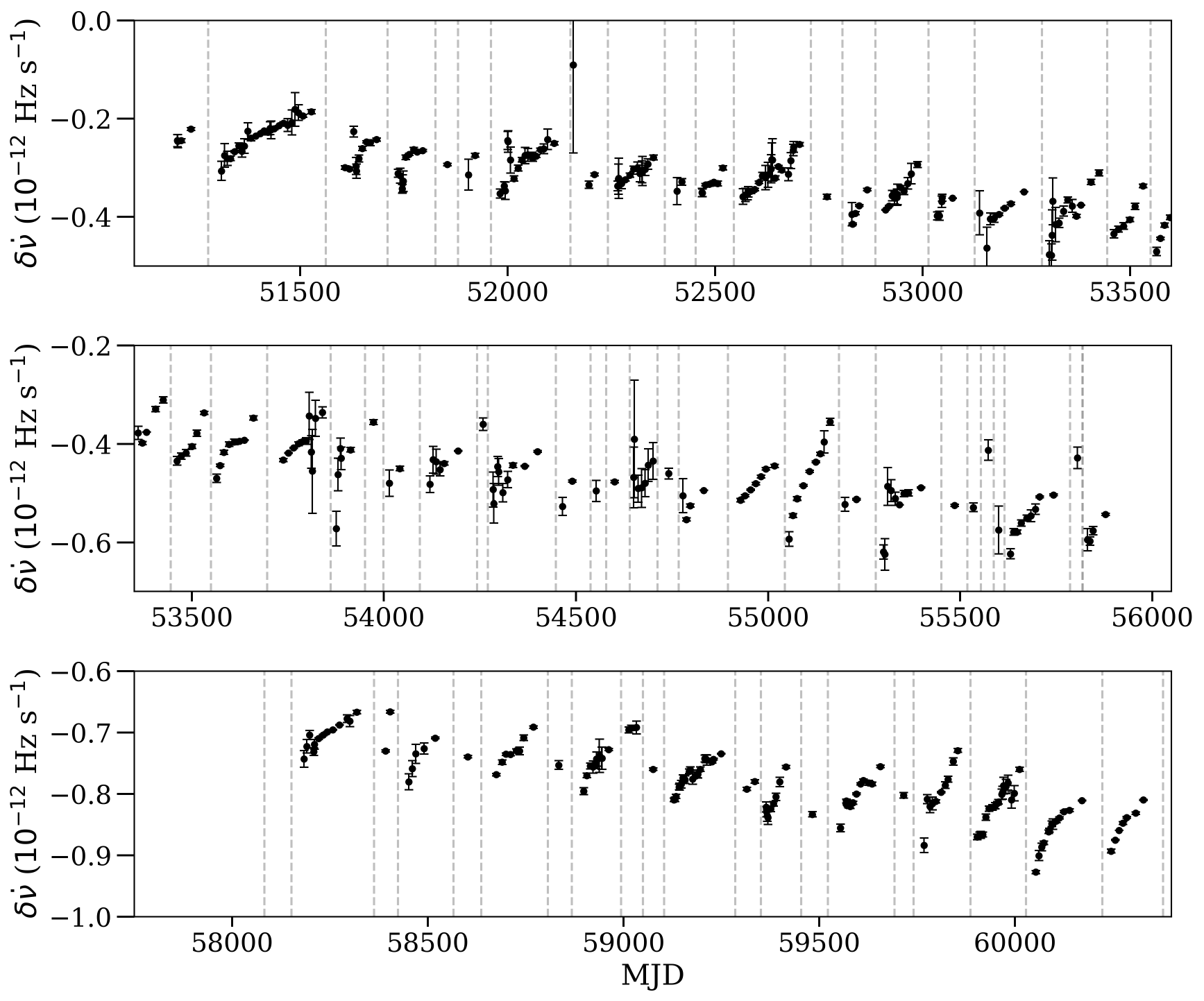}
    \caption{Evolution of the spin-down rate. For visual purposes, the quantity $\delta\dot\nu=\dot\nu+199\times10^{-12}$ was plotted. The full dataset was divided as in Fig. \ref{fig:F1-resids2}. The dashed vertical lines mark the epochs of the glitches.}
    \label{fig: spin-down}
\end{figure}

We present the evolution of $\dot\nu$ in Fig.~\ref{fig: spin-down}. The evolution shows a global negative slope, which is caused by repeated large glitches. The slope becomes positive when the effects of the glitches are removed.

\end{document}